\documentclass[a4paper,fleqn]{cas-dc}

\usepackage[numbers]{natbib}

\def\tsc#1{\csdef{#1}{\textsc{\lowercase{#1}}\xspace}}
\tsc{WGM}
\tsc{QE}
\tsc{EP}
\tsc{PMS}
\tsc{BEC}
\tsc{DE}

\usepackage{amsmath,amssymb,amsfonts}
\usepackage{algorithmic}
\usepackage{graphicx}
\usepackage{algorithm,algorithmic}
\usepackage{hyperref}
\hypersetup{hidelinks=true}
\usepackage{textcomp}
\usepackage{algorithmic}
\usepackage[table,xcdraw]{xcolor}
\usepackage{xcolor}
\usepackage{multirow}
\usepackage{amsmath}
\usepackage{booktabs}
\usepackage{bbm} 
\usepackage{bbding}
\usepackage{utfsym}
\usepackage{colortbl}
\usepackage{tabularx}

\usepackage{enumitem}
\setlist{labelindent=\parindent, leftmargin=*}
\usepackage{array}

\usepackage{ragged2e}
\usepackage{pifont}
 \usepackage{amssymb}

\begin{document}
\let\WriteBookmarks\relax
\def\floatpagepagefraction{1}
\def\textpagefraction{.001}
\shorttitle{}
\shortauthors{Zixian Su and Jingwei Guo, et~al.}

\title [mode = title]{Navigating Distribution Shifts in  Medical Image Analysis: A Survey}                      


\author[1]{Zixian Su}[orcid=0000-0002-1750-1501]
\fnmark[1] 
\ead{zixian12138@163.com}

\author[2]{Jingwei Guo}[orcid=0000-0003-4336-9863]
\fnmark[1] 
\ead{jingweiguo19@outlook.com}

\author[3]{Xi Yang}[orcid=0009-0007-9693-7963]
\cormark[1] 
\ead{xi.yang01@xjtlu.edu.cn}

\author[3]{Qiufeng Wang}[orcid=0000-0002-0918-4606]
\ead{qiufeng.wang@xjtlu.edu.cn}

\author[4]{Frans Coenen}[orcid=0000-0003-1026-6649]
\ead{coenen@liverpool.ac.uk}

\author[5,6]{Amir Hussain}[orcid=0000-0002-8080-082X]
\ead{amir.hussain@phc.ox.ac.uk}

\author[7]{Kaizhu Huang}[orcid=0000-0002-3034-9639]
\cormark[1] 
\ead{kaizhu.huang@dukekunshan.edu.cn}


\affiliation[1]{organization={Life Simulation Research Center, Beijing Academy of Artificial Intelligence},
                city={Beijing},
                country={China}}

\affiliation[2]{organization={Electrical and Mathematical Sciences and Engineering Division, King Abdullah University of Science and Technology},
                city={Thuwal},
                country={Kingdom of Saudi Arabia}}

\affiliation[3]{organization={Department of Intelligent Science, School of Advanced Technology, Xi’an Jiaotong-Liverpool University},
                city={Suzhou},
                country={China}}

\affiliation[4]{organization={Computer Science, School of Computer Science and Informatics, University of Liverpool},
                city={Liverpool},
                country={UK}}

\affiliation[5]{organization={SDAIA-KFUPM Joint Research Centre for Artificial Intelligence, King Fahd University of Petroleum and Minerals},
                city={Dhahran},
                country={Kingdom of Saudi Arabia}}

\affiliation[6]{organization={Nuffield Department of Primary Care Health Sciences, University of Oxford},
                city={Oxford},
                country={UK}}

\affiliation[7]{organization={Digital Innovation Research Center, Electrical and Computer Engineering, Duke Kunshan University},
                city={Suzhou},
                country={China}}

\cortext[cor1]{Corresponding authors.}
\fntext[fn1]{Equal contribution.}
\fntext[fn2]{A preprint and has been submitted for consideration.}

\begin{abstract}
Medical Image Analysis (MedIA) has become indispensable in modern healthcare, enhancing clinical diagnostics and personalized treatment. Despite the remarkable advancements supported by deep learning (DL) technologies, their practical deployment faces challenges posed by distribution shifts, where models trained on specific datasets underperform on others from varying hospitals, or patient populations. To address this issue, researchers have been actively developing strategies to increase the adaptability of DL models, enabling their effective use in unfamiliar environments. This paper systematically reviews approaches that apply DL techniques to MedIA systems affected by distribution shifts. Rather than organizing existing methods by technical characteristics, we explicitly bridge real-world clinical constraints -- such as limited data accessibility, strict privacy requirements, and heterogeneous collaboration protocols -- with the technical paradigms able to address them. By establishing this connection between operational constraints and methodological evolution, we categorize existing works into Joint Training, Federated Learning, Fine-tuning, and Domain Generalization, each aligned with specific healthcare scenarios. Beyond this taxonomy, our empirical analysis suggests that, as domain information becomes progressively less accessible across these paradigms, performance improvements become increasingly constrained, and further uncovers a gradual shift in methodological focus from explicit distribution alignment toward uncertainty-aware modeling, ultimately pointing to the need for more deployability-aware design in real-world MedIA.
\end{abstract}

\begin{graphicalabstract}
\includegraphics[width=\textwidth]{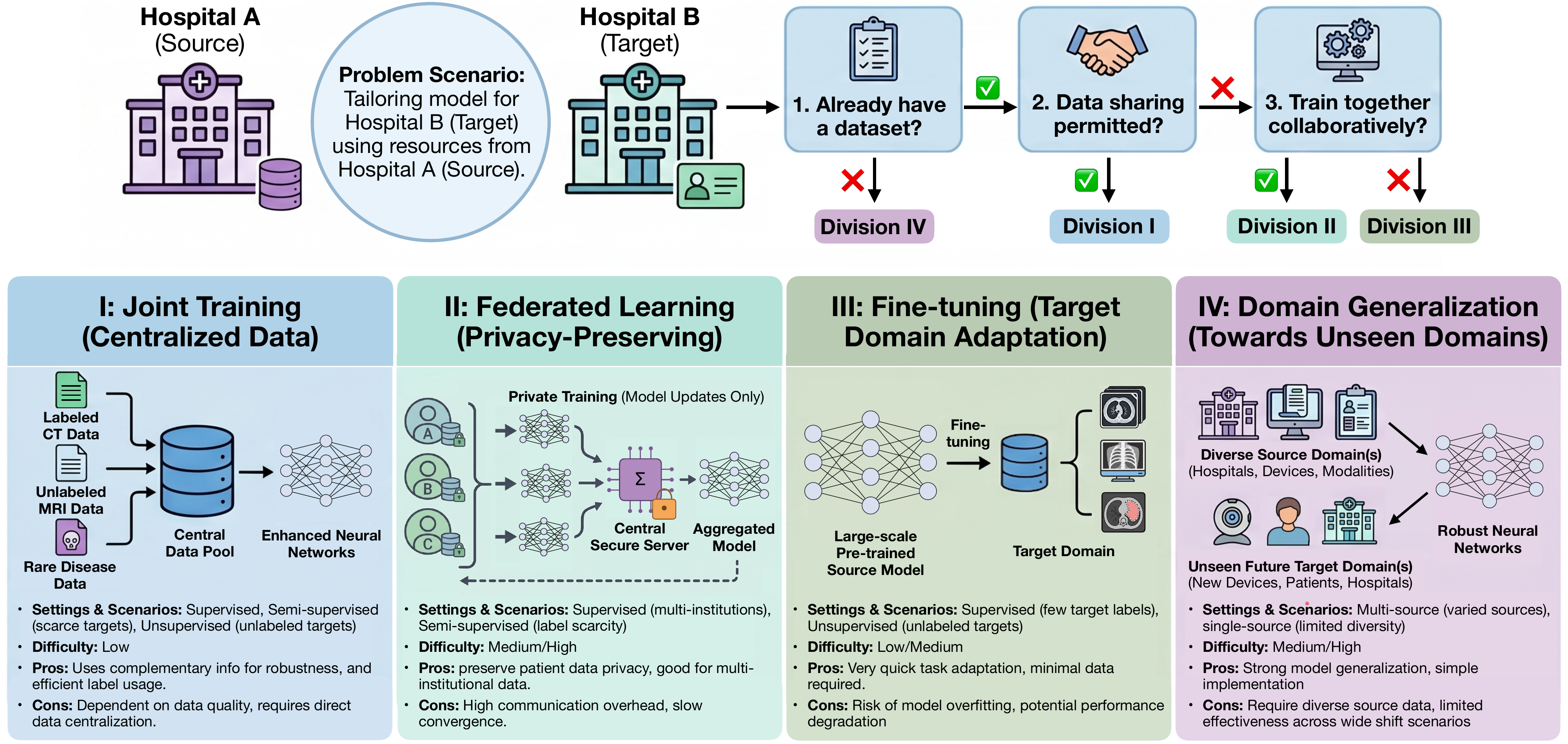}
\end{graphicalabstract}

\begin{highlights}
\item Introduce a constraint-aware taxonomy for distribution shifts in MedIA
\item Bridge clinical deployment scenarios with methodological paradigms
\item Analyze how domain information access shapes methodological evolution
\end{highlights}


\maketitle


\section{Introduction}\label{sec:introduction}
{M}{edical} image analysis (MedIA)~\cite{media,acmdataset} has become a cornerstone of modern healthcare, playing a critical role in enhancing diagnostics~\cite{diagnosis1,diagnosis2}, patient monitoring~\cite{monitoring}, and treatment planning~\cite{treatment}. With the advent of high-resolution imaging technologies and the increasing complexity of medical data, the application of advanced computational tools has become indispensable. Deep learning (DL) technologies~\cite{huang2019deep,unet,medvit,medsam,hassija2024interpreting}, in particular, have revolutionized MedIA by enabling automated and accurate analyses of medical images~\cite{DL1,DL2,acm2021covid,acm2024breastcancer,acm2023covid,acm2018knee,acm2021ultrasound,guo2016adaptive,zhang20262d,abbas2025xrdnet,arshad2025novel}. These technologies leverage large datasets to train models that can recognize patterns with a precision often surpassing human capabilities~\cite{malik2020comparison}. The integration of DL in MedIA not only speeds up diagnostic processes but also offers the potential for personalized healthcare through more accurate patient-specific assessments.

\begin{figure*}
  \centering
  \includegraphics[width=0.98\textwidth]{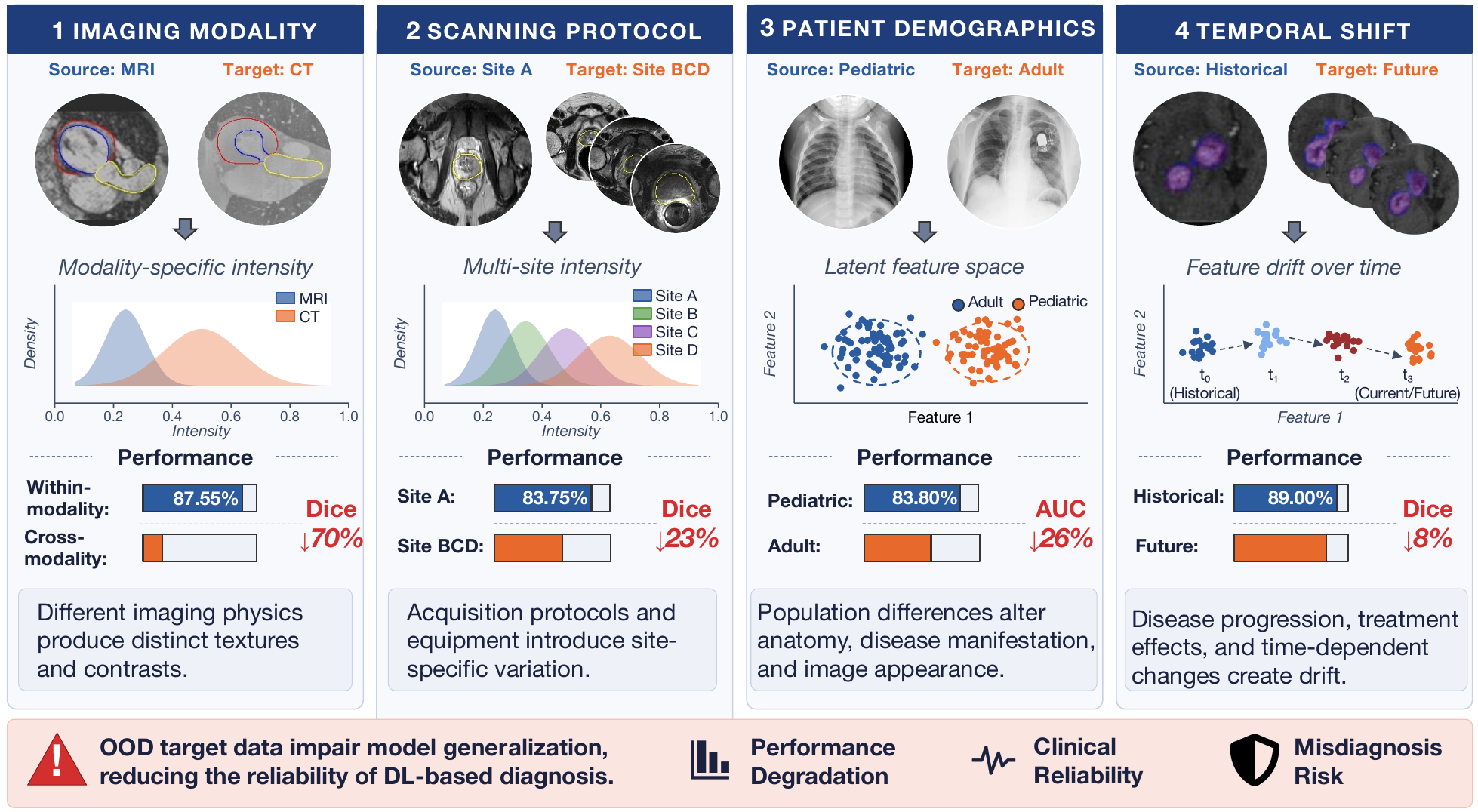}
  \caption{Illustration of medical imaging distribution shifts, showcasing from Imaging Modalities (Cardiac Substructure~\cite{MMWHS}), Scanning Protocols (Cross-Site Prostate~\cite{MDG4}), Patient Demographics  (Cross-Population Chest~\cite{kermany2018identifying,wang2017chestx,behera2025domain}), and Temporal Shifts 
  (Brain Tumor~\cite{jalalifar2020cascaded}).
  See details in  Seciton~\ref{sec:ds_in_media}.}
  \label{fig:banner_1}
\end{figure*}

However, the application of deep learning techniques in MedIA faces substantial challenges, primarily due to distributicon shifts. These shifts occur because the training data (known as source data) used to develop DL models often come from highly controlled environments or specific populations (see Figure~\ref{fig:banner_1}). When deployed in varied medical settings -- like different hospitals, population regions, and time periods -- these models encounter data that differ significantly in aspects such as imaging modalities~\cite{MMWHS}, scanning protocols~\cite{MDG4}, patient populations~\cite{Pediatric}, and temporal changes~\cite{temporal}. These variations expose the models to novel, out-of-distribution patterns (referred to as target data) that they have not been trained to recognize, impairing their ability to generalize effectively; this compromised performance in turn undermines the reliability and effectiveness of DL-based diagnostics.
Therefore, addressing these distribution shifts is crucial for the effect and reliable deployment of DL technologies in diverse medical environments.

To this end, this survey focuses on investigating DL-based MedIA under the challenges posed by distribution shifts. In recent years, the research community has actively developed strategies to enhance the adaptability and robustness of DL models. These strategies
aim to mitigate the impact of data distribution shifts across diverse medical settings~\cite{guan2021domain,yoon2023domain,tan2025towards}. In real-world healthcare, the successful deployment of DL technologies often encounters various operational constraints that directly leads to different data distribution shift scenarios. These constraints typically stem from several key factors:
\begin{itemize}
    \item\textbf{Data Accessibility:} This aspect 
    concerns the availability of comprehensive datasets for training DL models. 
    The breadth and quality of accessible data impacts how well a model can be trained to handle varied medical conditions, determining the difficulty level of managing the potential data distribution shifts.
    \item\textbf{Privacy Concerns:} Given the sensitive nature of medical data, privacy concerns~\cite{privacy} revolve around the protection of patient information.
    These considerations often limit the sharing of medical data among different healthcare institutions, creating data silos that exacerbate the potential data distribution shifts.
    \item\textbf{Collaborative Protocols:} Collaboration among healthcare institutions enables collective efforts to improve diagnostic models across diverse settings.
    By adhering to different protocols, various collaborative methods~\cite{Collaborative1,Collaborative2} have been developed while meeting specific requirements to alleviate the potential distribution shifts.
\end{itemize}

\begin{figure*}[t]
  \centering
\includegraphics[width=0.98\textwidth]{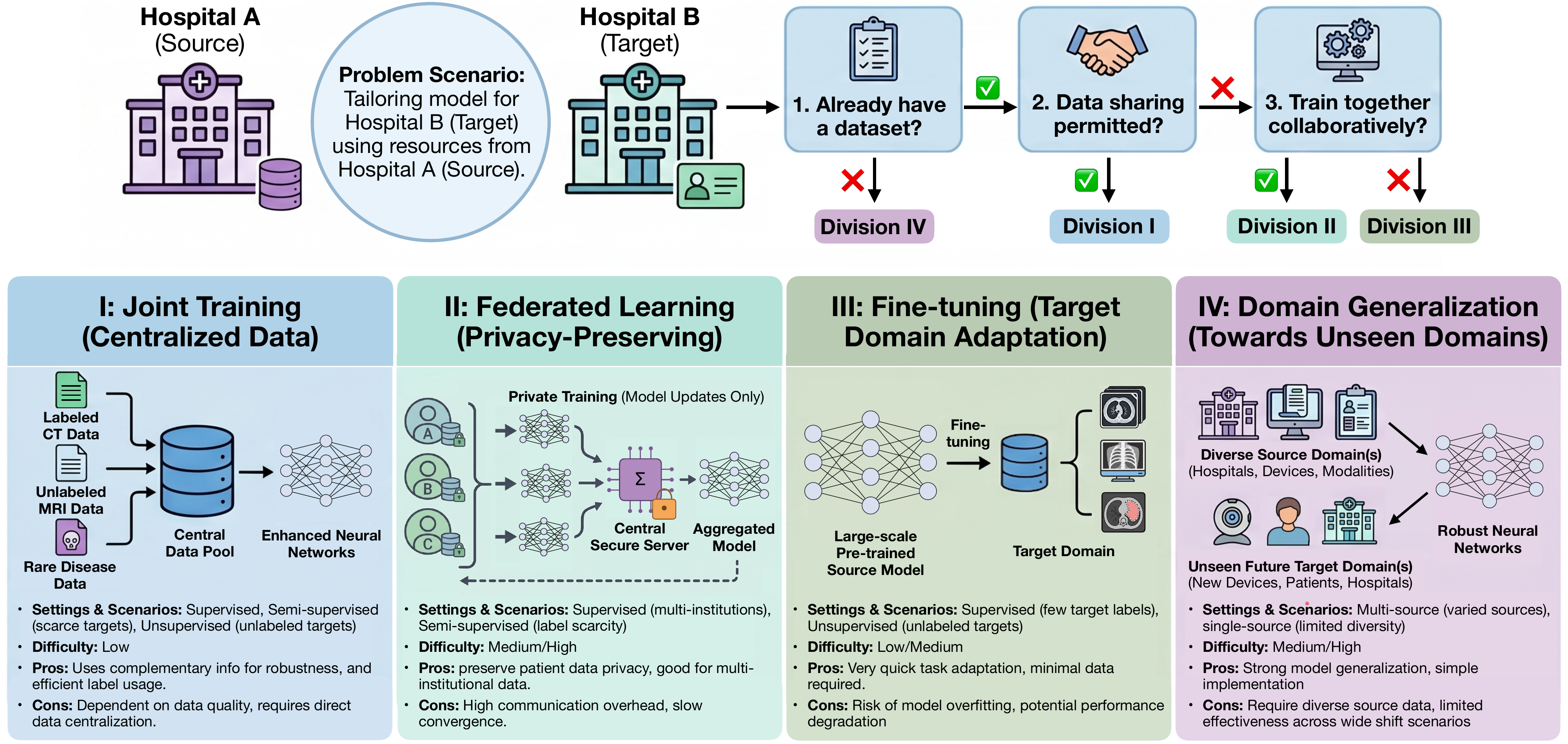}
  \caption{
 The diagram categorizes existing deep learning techniques into four main approaches,  each addressing real-world operational constraints including  Data Accessibility, Privacy Concerns, and Collaborative Protocols.  In Joint Training, hospitals collaborate by sharing data for model training. Federated Learning enables collaboration without direct data sharing, maintaining privacy. Fine-tuning adapts pre-trained models from one institution to another. Domain Generalization develops models that generalize across diverse settings, even without access to target data, to mitigate distribution shifts.
  }
  \label{fig:banner}
\end{figure*}

\noindent Building on these practical considerations, 
we categorize existing efforts to manage distribution shifts in MedIA into a hierarchy from simple to hard (see Figure~\ref{fig:banner}):
\begin{itemize}
    \item \textbf{Joint Training:} This approach is feasible when both the source and target data are accessible and there is no privacy concerns. This scenario often occurs when multiple health institutions agree to share their own data, facilitating joint model training~\cite{joint7,chen2019synergistic} and thereby enhancing model adaptability across diverse settings.
    \item \textbf{Federated Learning:} When multiple institutions seek to cooperate without exposing their distinct datasets due to privacy concerns, federated learning~\cite{acm2022federated,fedavg} offers a powerful solution. It 
    enables collaborative model improvements across different institutions by training models locally and aggregating the learned models without centralizing data storage. 
    \item \textbf{Fine-tuning:} 
    When synchronous collaborations are not allowed for addressing data distribution shifts with privacy concerns, fine-tuning~\cite{finetune1,TTA3} emerges as an effective remedy. This involves using a well pre-trained model and then fine-tuning it on new datasets to transfer learned knowledge to unfamiliar domains. 
    \item \textbf{Domain Generalization:} 
    When data from unseen domains that require model adaptation is inaccessible or unknown, training a model that is generalizable enough to withstand distribution shifts is essential~\cite{MDG5,SDG2}. This involves preparing for unforeseen challenges by developing models that can generalize from the data currently available for training to any potential new environments. 
\end{itemize}

In this survey, we present a nuanced understanding of how DL can be strategically deployed to address distribution shifts in MedIA. While a few existing surveys have explored distribution shifts in MedIA, they predominantly concentrate on the technical mechanics of existing approaches --- categorizing methods based on the degree of DL supervision~\cite{guan2021domain} or common MedIA workflows~\cite{yoon2023domain} -- or restrict their discussions to specific organs (e.g., heart, lung, brain)~\cite{acm2023nuerocrossmodality,li2023multi,kora2022transfer,yu2022transfer,kim2022transfer}. These conventional taxonomies often treat MedIA merely as an application domain, overlooking the real-world medical constraints that fundamentally shape these distribution shift scenarios.

To address this limitation, this survey is grounded in the practical, operational constraints faced by healthcare institutions. In doing so, it serves as a practical guide for medical professionals seeking to apply these methods to real clinical problems.
Furthermore,  by systematically analyzing empirical results across different paradigms, we highlight how these practical MedIA constraints drive algorithm design under distribution shifts: As explicit data alignment becomes unfeasible, research naturally pivots toward uncertainty-aware modeling and disentangled representation learning.
By linking technical trends to real-world constraints, our survey highlights that model performance is fundamentally bounded by domain information accessibility, and that the central challenge lies in balancing deployability and performance under such constraints. This perspective further points to a key direction: transforming real-world limitations into effective modeling priors, through which we aim to provide a principled roadmap for bridging algorithmic innovation and reliable clinical deployment.

\begin{figure*}[t]
  \centering
\includegraphics[width=0.98\textwidth]{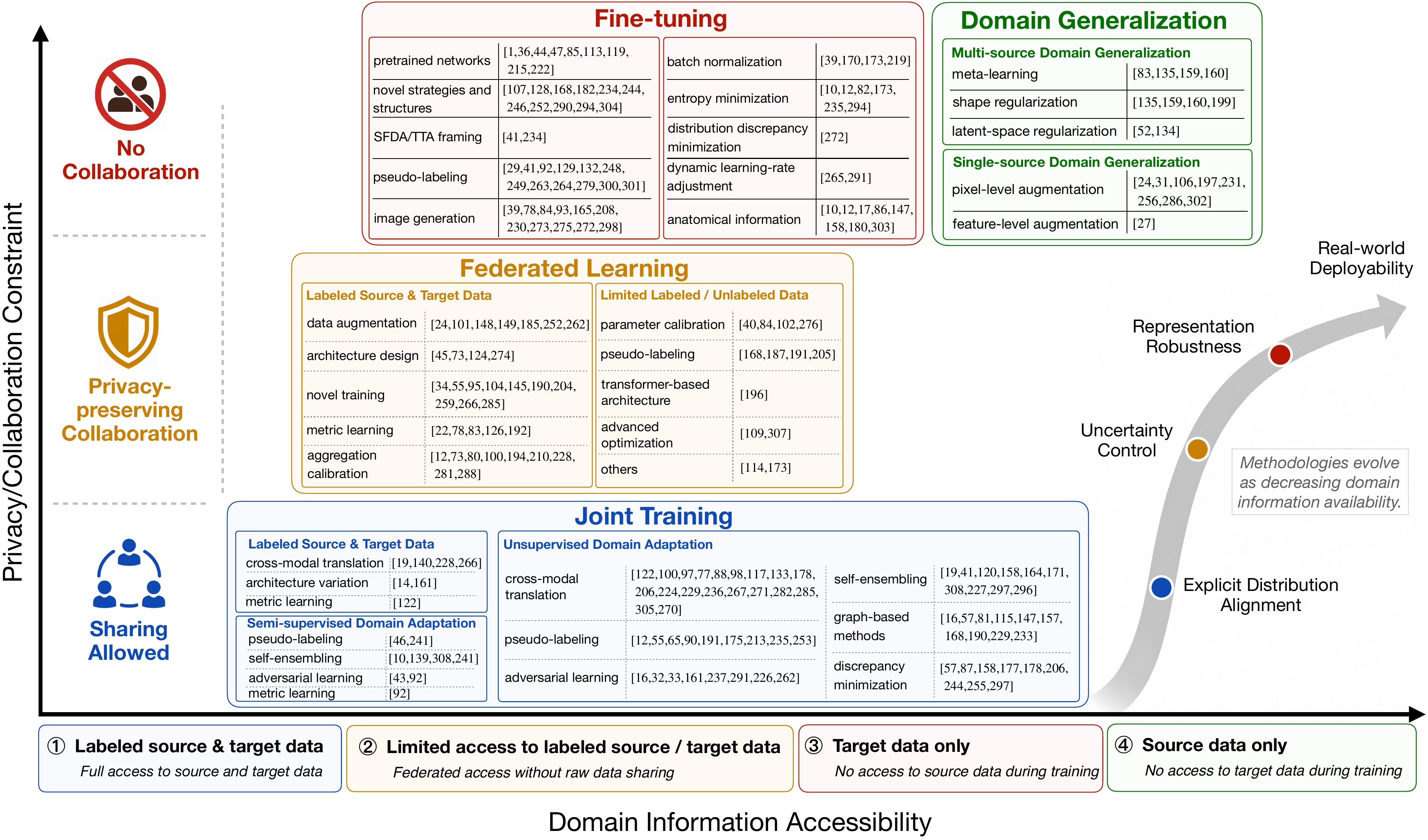}
\caption{\textbf{Constraint-aware taxonomy of distribution-shift learning in medical image analysis.} It organizes method families along privacy/collaboration constraints and domain information accessibility, showing how reduced data access and collaboration drive a transition from explicit alignment to uncertainty-aware, robust, and deployable learning.}
  \label{fig:detailed_taxonomy}
\end{figure*}

\section{Medical Distribution Shifts}\label{sec:ds_in_media}
The efficacy of DL models largely hinges on the assumption that the training and testing data are independently and identically distributed (i.i.d). However, this assumption often does not hold in the complex and diverse environment of clinical practice. The inherent heterogeneity in medical imaging, arising from different modalities, varying protocols, diverse patient demographics, and temporal shifts, introduces significant distribution shifts. In the following, we provide a concise illustration (also see Figure~\ref{fig:banner_1}): 

\begin{itemize}
    \item \textbf{Imaging Modalities}: Medical imaging encompasses a range of modalities, such as Magnetic Resonance Imaging (MRI), Computed Tomography (CT), X-rays, and Ultrasound, each producing images with unique characteristics. A model trained on data from one modality might not generalize well to another, given the inherent differences in image textures, contrasts, and anatomical representations. 
    
    \item \textbf{Scanning Protocols}: Even within the same modality, images can vary based on the imaging protocols and equipment used. Factors such as magnetic field strength in MRI, radiation dose in CT, and ultrasound machine settings can introduce significant variations in the images. 

    \item \textbf{Patient Demographics}: Differences in patient populations, such as age, gender, and ethnicity, as well as variations in disease manifestations, can lead to substantial differences in imaging data. For instance, pediatric images are markedly different from adult images. 

    \item \textbf{Temporal Shifts}: Longitudinal data collected over extended periods often encounter shifts due to physiological states change, treatment impact, and the progression of diseases. As a result, models trained on historical data may not perform optimally on current or future data. 
\end{itemize}

\section{Categorization and Frameworks}

This section introduces our categorization rationale for addressing distribution shifts in MedIA. As discussed in the preceding section, real-world deployment often begins from an operational problem: when a target institution, such as Hospital B, seeks to collaborate with a source institution, e.g., Hospital A, what information and collaboration plans are actually available? 
This practical setting can be decomposed into three questions: whether target-domain data exist, whether inter-institutional data sharing is permitted, and whether collaborative training is feasible without pooling data.
These questions determine not only which data are accessible, but also which learning paradigms are feasible under clinical, legal, and organizational constraints.

Building on this scenario-driven view, we organize existing DL methods using a two-dimensional taxonomy, as illustrated in Figure~\ref{fig:detailed_taxonomy}. The first dimension is \textbf{Domain Information Accessibility}, which describes what source and target information is available during training or adaptation. This dimension ranges from full access to labeled source and target data, to limited labeled or unlabeled access, target-data-only settings, and source-data-only settings. The second dimension is \textbf{Privacy Constraint}, which describes the degree to which institutions are allowed to share data or collaborate, ranging from direct data sharing, to privacy-preserving collaboration, and finally to no collaboration. 
To complement this high-level map, Table~\ref{tab:conditions-table} offers a more fine-grained characterization of the setting-level assumptions underlying our constraint-aware framework. Rather than serving as a taxonomy of methods, the table explicitly delineates the information accessibility and operational constraints associated with each setting.
Under this framework, Joint Training, Federated Learning, Fine-tuning, and Domain Generalization are not merely algorithmic families, but correspond to different real-world deployment scenarios. Joint Training assumes the richest access to source and target data and thus enables direct distribution alignment. Federated Learning addresses collaborative learning when raw data cannot be shared. Fine-tuning covers settings where adaptation must rely on limited target information or a pretrained source model. Domain Generalization represents the most restrictive scenario, where models must generalize to unseen target domains without access to target data during training.

\begin{table*}[t]
\caption{Constraint-aware summary of the our taxonomy, clarifying what information is available, unavailable, or constrained under each paradigm for MeddIA under distribution shifts.}\label{tab:conditions-table}
\centering
\resizebox{0.98\textwidth}{!}{
\begin{tabular}{>{\centering\arraybackslash}m{2.0cm}
                >{\centering\arraybackslash}m{1.9cm}
                >{\centering\arraybackslash}m{1.8cm}
                >{\centering\arraybackslash}m{1.5cm}
                >{\centering\arraybackslash}m{1.5cm}
                >{\centering\arraybackslash}m{1.9cm}
                >{\centering\arraybackslash}m{1.6cm}
                >{\centering\arraybackslash}m{1.6cm}
                >{\centering\arraybackslash}m{1.6cm}}
\toprule
Methods & Settings & Source Data Access & Source Labels & Target Labels & Target Adaptation & Multi-Source & Data Privacy & Model Privacy \\
\midrule
\multirow{3}{*}[-12pt]{Joint Training}
 & Supervised      & \checkmark & \checkmark & Full      & \checkmark & Optional & \ding{55} & \ding{55} \\
\cmidrule{2-9}
 & Semi-supervised & \checkmark & \checkmark & Partial   & \checkmark & Optional & \ding{55} & \ding{55} \\
\cmidrule{2-9}
 & Unsupervised    & \checkmark & \checkmark & \ding{55} & \checkmark & Optional & \ding{55} & \ding{55} \\
\midrule
\multirow{2}{*}[-6pt]{\shortstack{Federated\\ Learning}}
 & Supervised      & Decentralized & \checkmark & Full    & Collaborative & \checkmark & \checkmark & \checkmark \\
\cmidrule{2-9}
 & Semi-supervised & Decentralized & Partial & Partial & Collaborative & \checkmark & \checkmark & \checkmark \\
\midrule
\multirow{2}{*}[-6pt]{Fine-tuning}
 & Supervised      & \ding{55} & \ding{55} & Full      & \checkmark & Optional & \checkmark & \ding{55} \\
\cmidrule{2-9}
 & Unsupervised    & \ding{55} & \ding{55} & \ding{55} & \checkmark & Optional & \checkmark & \ding{55} \\
\midrule
\multirow{2}{*}[-6pt]{\shortstack{Domain\\ Generalization}}
 & Multi-source    & \checkmark & \checkmark & \ding{55} & \ding{55} & \checkmark & \checkmark & \ding{55} \\
\cmidrule{2-9}
 & Single-source   & \checkmark & \checkmark & \ding{55} & \ding{55} & \ding{55} & \checkmark & \ding{55} \\
\bottomrule
\end{tabular}}
\end{table*}

This taxonomy also reveals a broader methodological evolution. When source and target data are both accessible, methods can directly reduce domain discrepancy through \textit{explicit distribution alignment}. As target information becomes limited and raw data sharing becomes restricted, methods increasingly rely on \textit{uncertainty control} to improve adaptation considering incomplete supervision and noisy target signals. Under even stronger constraints, especially when target-domain data are unavailable, the emphasis shifts toward \textit{representation robustness}, where models are encouraged to learn features that remain stable across heterogeneous domains. Ultimately, the most clinically relevant goal is \textit{real-world deployability}: developing models that can remain reliable across institutions, scanners, populations, and temporal changes despite limited domain information and collaboration constraints.

After establishing this constraint-aware taxonomy, we further provide a technical organization to help readers quickly identify methods relevant to their own deployment setting from three technical aspects (see Table~\ref{tab:taxonomy_new}):

\begin{itemize}
\item \textbf{Data Engineering}: Methods that increase the model's exposure to diverse data scenarios through data augmentation, selection, synthesis, translation, or domain simulation.
\item \textbf{Model Design}: Methods that modify network architectures, representation spaces, normalization modules, or structural priors to improve adaptability and robustness under distribution shifts.
\item \textbf{Optimization Strategy}: Methods that adjust the learning objective, training schedule, pseudo-labeling process, uncertainty estimation, or regularization strategy to better learn from shifted and heterogeneous data.
\end{itemize}

\section{Preliminaries}
A domain $\mathcal{D}$ is a joint distribution $p(x,y)$ defined on the input-output space $\mathcal{X} \times \mathcal{Y}$, where random variables $x \in \mathcal{X}$ and $y \in \mathcal{Y}$ denote the input data and the output label, respectively.
We typically deal with two distinct datasets, known as the \textbf{source} and \textbf{target} domains. 
The \textbf{Source Domain} $\mathcal{D}_{s} = \{(x,y) \sim p_s(x,y)\}$.
comprises medical images $x$ such as X-rays or MRI scans, each paired with a label $y$ that might be categorical information regarding disease diagnosis or the segmentation mask. The \textbf{Target Domain} $\mathcal{D}_{t} = \{(x,y) \sim p_t(x,y)\} $ originates from a different but related distribution to that of the source. For instance, they might come from different medical imaging devices or patient populations. Note that for both source and target distributions, $p_s(x,y) = p_s(x)p_s(y| x)$ and $p_t(x,y) = p_t(x)p_t(y| x)$. We take the standard covariate shift assumption as \textbf{Distribution Shift}, i.e., 
$p_s(y| x) = p_t(y|x)$ and $p_s(x) \neq p_t(x)$.
In this situation, the model  $q_{{\theta}}(y|x)$ solely trained on the source domain cannot well represent the true, domain-invariant distribution $p(y|x)$. Therefore, a variety of research concentrates on adjusting $q_{{\theta}}(y|x)$ to maximize its predictive performance on the target distribution.

\section{Joint Training}\label{sec:joint_training}
\begin{figure}
  \centering
\includegraphics[width=0.45\textwidth]{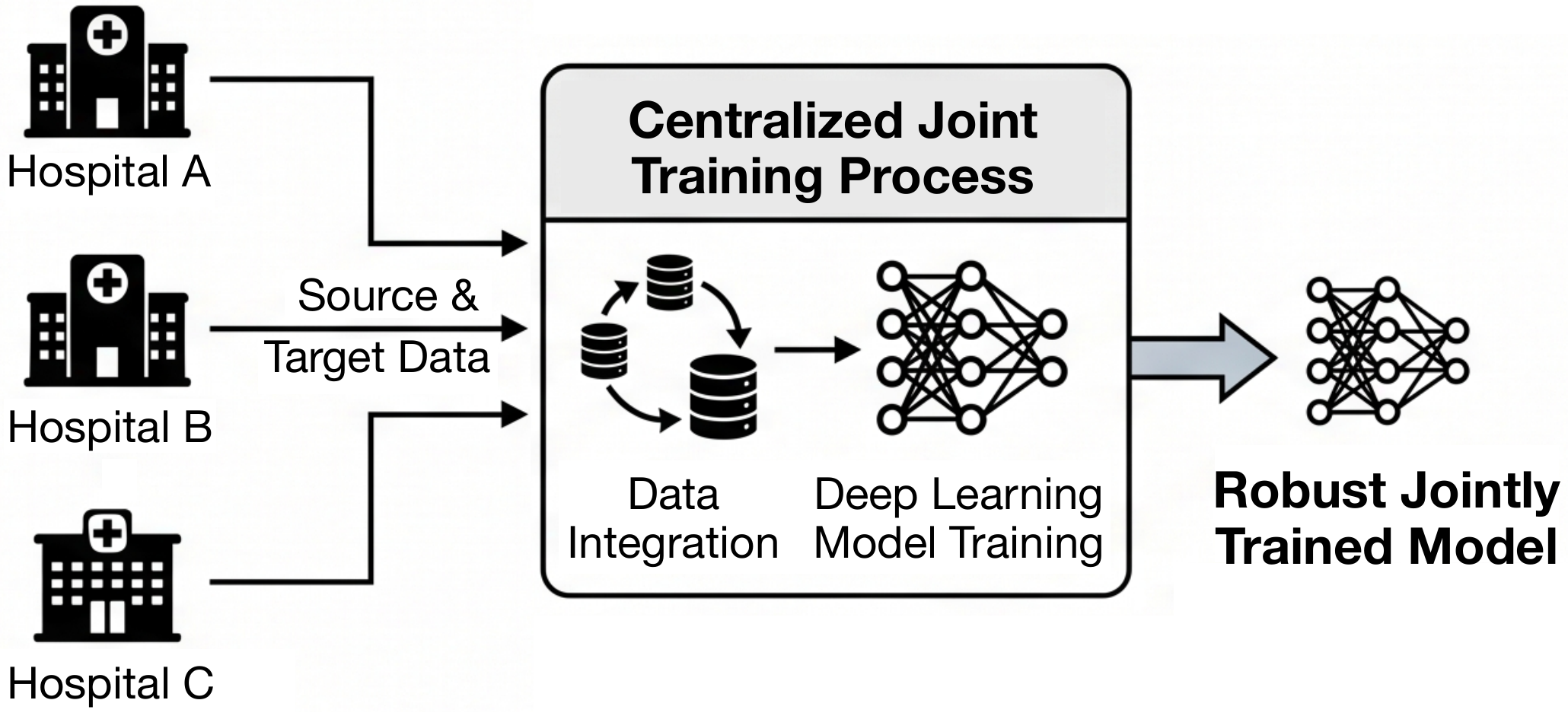}
  \caption{Joint Training in MedIA: It enables data sharing among healthcare institutions without privacy constraints, while integrating source and target data to improve model adaptability across diverse medical settings.}
\label{fig:banner_jt}
\end{figure}

Joint Training is a crucial domain adaptation strategy in MedIA, particularly effective when target data is freely accessible and privacy concerns are minimal. This method excels in environments where healthcare institutions can collaboratively share data, creating the ideal conditions for joint model training. Such collaboration significantly enhances the adaptability of models across varied medical settings by integrating both source and target data, as shown in Fig.~\ref{fig:banner_jt}. Typically, the source dataset is fully-labeled, whereas the target dataset often exhibits varying labeling rates due to changes in medical scenarios, introducing complexities to DL model training. In response, a variety of joint training strategies have emerged, each designed to address the specific challenges posed by fluctuating label availability on target data. These methods are categorized based on the level of target supervision, ranging from Supervised to Semi-supervised, and Unsupervised Joint Training.

\subsection{Supervised Joint Training}\label{sec:supervised_joint_training}
Supervised Joint Training is a domain adaptation strategy where models are concurrently trained on both the source and target domain data, leveraging labeled data from both to enhance performance despite domain shifts. This method is particularly valuable when the target domain has significantly less labeled data than the source, as relying solely on target data would yield inadequate model performance. In Supervised Joint Training, the strategy involves integrating different modalities or varying views of data, and often includes synthesizing data to mitigate the data shortage problem in the target domain. These methods effectively utilize the structural and distributional characteristics of data from both domains, making full use of all available labeled data to effectively bridge the gap between the source and target domains.

\newcommand{\tabitem}{\makebox[1.2em][l]{$\scriptstyle\bullet$}}
\renewcommand{\tabularxcolumn}[1]{m{#1}}
\newcolumntype{S}{>{\hsize=0.95\hsize\raggedright\arraybackslash}X}
\newcolumntype{M}{>{\hsize=1.0\hsize\raggedright\arraybackslash}X}
\newcolumntype{W}{>{\hsize=1.05\hsize\raggedright\arraybackslash}X}

\begin{table*}[t]
\centering
\caption{Methodological Taxonomy Across Data Engineering, Model Design, and Optimization Strategy Aspects.}
\label{tab:taxonomy_new}
\setlength{\tabcolsep}{1pt}
\resizebox{1.0\textwidth}{!}{
\setlength{\parskip}{3pt}        
\renewcommand{\arraystretch}{1.5} 
\begin{tabularx}{1.32\textwidth}{>{\bfseries\raggedright\arraybackslash}m{2.2cm} S M W}
\toprule
 & \multicolumn{1}{c}{\textbf{Data Engineering}} & \multicolumn{1}{c}{\textbf{Model Design}} & \multicolumn{1}{c}{\textbf{Optimization Strategy}} \\ 
\midrule

{Joint Training} & 
\tabitem Cross-modal Translation (Sec.~\ref{sec:cross_model_traslation_1} \&~\ref{sec:cross_model_translation_2}) \par
\tabitem Pseudo-labeling (Sec.~\ref{sec:pseudo_label} \&~\ref{sec:plabel_v2}) & 
\tabitem Architecture Variations (Sec.~\ref{sec:arch_var_1}) \par
\tabitem Self-ensembling (Sec.~\ref{sec:self_ensemble} \&~\ref{sec:selfensemble_2}) \par
\tabitem Adversarial Learning (Sec.~\ref{sec:adv_learning_1} \&~\ref{sec:adlearn_v2}) \par
\tabitem Novel Training Strategies (Sec.~\ref{sec:novel_training_strag_1}) \par
\tabitem Graph-based Methods (Sec.~\ref{sec:graph_based_1}) & 
\tabitem Metric Learning (Sec.~\ref{sec:metric_learning_v1} \&~\ref{sec:metric_learning_v2_new}) \par
\tabitem Discrepancies Minimization (Sec.~\ref{sec:sta_dis_mini}) \\ 
\midrule

{Federated Learning} & 
\tabitem Data Augmentation (Sec.~\ref{sec:data_aug_1}) \par
\tabitem Pseudo-labeling (Sec.~\ref{sec:peseudo_labeling_v3}) & 
\tabitem Novel Architecture Design (Sec.~\ref{sec:novel_arch_design_v1}) \par
\tabitem Novel Training Strategies (Sec.~\ref{sec:novel_train_stra_v2} \&~\ref{sec:novel_training_stratg_v3}) \par
\tabitem Transformer-based Architecture (Sec.~\ref{sec:transformer_arch_v2}) & 
\tabitem Metric Learning (Sec.~\ref{sec:metric_learning_v2}) \par
\tabitem Aggregation Weight Calibration (Sec.~\ref{sec:agg_weig_calb}) \par
\tabitem Parameter Calibration (Sec.~\ref{sec:param_clab}) \par
\tabitem Advanced Optimization Strategies (Sec.~\ref{sec:advance_opt_stratg}) \\ 
\midrule

{Fine-tuning} & 
\tabitem Pseudo-labeling (Sec.~\ref{sec:pesu-label-v3}) \par
\tabitem Image Generation (Sec.~\ref{sec:image_cap}) & 
\tabitem Novel Pipelines (Sec.~\ref{sec:Novel Strategies and Structures} \&~\ref{sec:Novel_Strategies_and_Structures_v2}) \par
\tabitem Batch Normalization (Sec.~\ref{sec:batch_normalization}) & 
\tabitem Entropy Minimization (Sec.~\ref{sec:entropy_min}) \par
\tabitem Discrepancies Minimization (Sec.~\ref{sec:distribution_discrep_mini}) \par
\tabitem Dynamic Learning Rates Adjustment (Sec.~\ref{sec:Dynamic Adjustment of Learning Rates}) \par
\tabitem Anatomical Information (Sec.~\ref{sec:anato_inform}) \\ 
\midrule

{Domain Generalization} & 
\tabitem Pixel-level Augmentation (Sec.~\ref{sec:Pixel-level Augmentation}) \par
\tabitem Feature-level Augmentation (Sec.~\ref{sec:Feature-level Augmentation}) & 
\tabitem Meta-learning (Sec.~\ref{sec:meta_learning_v2}) & 
\tabitem Shape-based Regularization (Sec.~\ref{sec:Shape-based Regularization}) \par
\tabitem Latent Space Regularization (Sec.~\ref{sec:Latent Space Regularization}) \\ 
\bottomrule
\end{tabularx}
}
\end{table*}


\subsubsection{Cross-modal Translation}\label{sec:cross_model_traslation_1}
Cross-modal translation plays a pivotal role in addressing the challenge of integrating data from diverse imaging modalities, which often exhibit distinct   intensity and texture characteristics. This technique facilitates the conversion of data between modalities, such as from MRI to CT images, enabling the use of a unified dataset for training despite the inherent discrepancies. By synthesizing data from one modality in the form that resembles another, cross-modal translation helps to overcome the shortage data problem and enhances the robustness of the training process. Specifically, Generative Adversarial Networks have proven to be particularly effective for cross-modal translation~\cite{joint4,joint5,semi2,joint10} by creating high-quality synthetic images that maintain the domain-specific characteristics of the target modality. For example, the shape-consistency approach~\cite{joint4} leverages GANs for volume-to-volume translation, ensuring that the structural integrity of medical images is preserved across modalities. 

\subsubsection{Architecture Variations}\label{sec:arch_var_1}
Novel architectural designs are crucial for addressing domain adaptation challenges. For instance, the domain-adaptive two-stream U-Net, applied for electron microscopy image segmentation~\cite{joint1}, features a dual-stream architecture that supports selective weight sharing between source and target domains. This design enhances adaptability by allowing the model to fine-tune its responses to the unique characteristics of each domain. Similarly, the Multi-Site Network (MS-Net) for cross-site prostate segmentation~\cite{joint7} incorporates Domain-Specific Batch Normalization (DSBN). DSBN effectively manages inter-site variability by providing distinct feature normalization for each site, ensuring that the model remains robust across diverse MRI datasets.

\subsubsection{Metric Learning}\label{sec:metric_learning_v1}
Metric Learning has proven instrumental in maintaining high generalization performance across different data domains. A notable application is demonstrated in~\cite{joint2}, where metric learning is employed to enhance domain adaptation for Wireless Capsule Endoscopy. This approach utilizes triplet loss, a form of contrastive learning, which effectively minimizes the distance between embeddings of samples with the same labels from different domains while maximizing the distance between samples with different labels from the same domains. By doing so, it ensures that the model can accurately interpret and classify medical images regardless of the specific device version.

\subsection{Semi-supervised Joint Training}\label{sec:semisupervised_joint_training}
Semi-supervised Joint Training, also referred to as Semi-Supervised Domain Adaptation (SSDA), is a cutting-edge machine learning strategy aimed at transferring knowledge from a well-labeled source domain to a target domain with scarce labels. This approach is vital in situations where acquiring comprehensive labels for the target domain is impractical due to cost or time constraints. The primary challenges of SSDA include maximizing the utility of limited labeled data and a larger volume of unlabeled data in the target domain, as well as mitigating distribution discrepancies between the domains.

\subsubsection{Pseudo-labeling}\label{sec:pseudo_label}
Pseudo-labeling is a powerful technique in semi-supervised joint training that leverages large volumes of unlabeled data to enhance model training. This method involves generating artificial labels for unlabeled data based on the most confident predictions of the model, thereby expanding the training dataset effectively. 
In domain adaptation, its effectiveness hinges on prioritizing model's predictions exceeding a defined threshold that are more likely to be accurate:
\begin{equation}
\hat{y}_u = \begin{cases} 
\arg\max q_\theta(x_u), & \text{if } \max q_\theta(x_u) > \tau \\
\text{ignore}, & \text{otherwise}
\end{cases}
\end{equation}
where \( x_u \) represents an unlabeled sample from the target domain, $q_\theta(x_u)$ denotes the model's probabilistic predictions, \( \hat{y}_u \) is the pseudo-label assigned to \( x_u \), and \( \tau \) is a threshold defining the confidence level above which the labels are considered reliable.
The loss is computed as:
\begin{equation}
\mathcal{L}_{u} = \sum_{x_u \in \mathcal{X}_u} \mathbb{I}\left(\max q_\theta(x_u) > \tau \right) \cdot \mathcal{L}(\hat{y}_u, q_\theta(x_u))
\end{equation}
where \( \mathbb{I}(\cdot) \) is an indicator function that selects high-confidence samples.
The total training objective combines the loss from labeled data and high-confidence pseudo-labeled data:
\begin{equation}
\mathcal{L}_{total} = \mathcal{L}_{l}( \mathcal{Y}_l, q_\theta(\mathcal{X}_l)) + \beta \mathcal{L}_{u}(\mathcal{\hat{Y}}_{u}, q_{\theta}(\mathcal{X}_{u}))
\end{equation}
where \( \beta \) is a balancing factor between the source and pseudo-labeled losses.

To mitigate the risk of error propagation, which can occur if incorrect labels are used for training, enhancements are made to ensure the quality of these pseudo-labels. For example, \cite{semi5} employs transformation-invariant, highly-confident predictions in the target dataset for self-training purposes, ensuring that the model is less likely to learn from noisy, less reliable labels. Meanwhile, \cite{semi7} enhances the robustness of pseudo-labeling by calculating the variance between the original image and its Fourier-transformed counterpart, providing a more stable basis for generating reliable pseudo-labels. These strategies significantly improve the utility of pseudo-labeling, making it a vital tool for utilizing unlabeled data in domain adaptation.

\subsubsection{Self-ensembling}\label{sec:self_ensemble}
Self-ensembling is an advanced learning strategy that effectively exploits both labeled and unlabeled data with the consistency between models. This method trains multiple versions of a model, each subjected to distinct input perturbations, and employs consistency regularization to ensure uniform predictions across these variations. 
Typically, this technique under SSDA setting is implemented via a ``teacher-student'' model, where a stable, pre-trained ``teacher'' model $q_{\theta}^{tc}$ guides a less-trained ``student'' model $q_{\theta}^{st}$. The overall training process is governed by two primary loss functions: 
\begin{align}
\nonumber
\mathcal{L}_{{total}} = \mathcal{L}_{l}(\mathcal{Y}_l, q_{\theta}^{tc}(\mathcal{X}_l)) + \lambda \mathcal{L}_{{cons}}(q_{\theta}^{tc}(\mathcal{X}_u)), q_{\theta}^{st}(\mathcal{X}_u^{aug}))
\end{align}

\noindent where supervised loss \(\mathcal{L}_{l}\) is for measuring discrepancies in the teacher's predictions on labeled input $q_{\theta}^{tc}(\mathcal{X}_l)$, and consistency loss \(\mathcal{L}_{\text{cons}}\)  is for aligning the teacher’s predictions on unlabeled input  $q_{\theta}^{tc}(\mathcal{X}_u)$ with the student’s on perturbed inputs $q_{\theta}^{st}(\mathcal{X}_u^{aug})$. The two loss functions are balanced by a regularization parameter \(\lambda\).

Following this framework, \cite{semi1,semi8} introduce this techniques into achieveing semi-supervised domain adaptation. The core innovation of these frameworks lies in the strategic employment of dual-teacher models: one teacher model enhances intra-domain knowledge via self-ensembling techniques, while the other facilitates inter-domain knowledge transfer using image translation models such as CycleGAN~\cite{zhu2017unpaired}. This approach leverages the consistency of model outputs across different views of the same data, enhancing the model's ability to generalize across diverse scenarios.

\subsubsection{Adversarial Learning}\label{sec:adv_learning_1}
Adversarial leaning is fundamentally embodied by Generative Adversarial Networks (GANs)~\cite{acm2022GAN}.
\noindent
It can be conceptualized as a game between two players:
\begin{equation}
\begin{split}
\min_G \max_D V(D, G) &=  \mathbb{E}_{x \sim p_{\text{data}}(x)}[\log D(x)] \\
& + \mathbb{E}_{z \sim p_z(z)}[\log (1 - D(G(z)))]
\end{split}
\end{equation}

\noindent where generator (G) aims to produce data that is indistinguishable from real data by transforming input noise \( z \), sampled from a noise distribution \( p_z(z) \), and discriminator (D) aims to correctly classify real data \( x \) and generated data \( G(z) \). Real data \( x \) is sampled from the true data distribution \( p_{\text{data}}(x) \). This interaction forms a min-max game where the generator seeks to deceive the discriminator into accepting its outputs as real, while the discriminator improves at identifying the differences between real and generated data. Through this adversarial process, $G$ refines its outputs to reduce discrepancies, indirectly generating domain-invariant features for domain adaptation.
As one notable method, COVID-DA~\cite{semi6} is designed to distinguish between closely related conditions such as pneumonia and COVID-19, particularly when labeled data is scarce. This method uses a unique classifier separation scheme along with an adversarial network to overcome the task difference and domain discrepancy simultaneously.

\subsubsection{Novel Training Strategies}\label{sec:novel_training_strag_1}

\cite{semi3}~explores the richness of multi-modal data through a novel asymmetric co-training approach.  By segmenting the learning process into two distinct components that each addresses specific aspects of domain adaptation and semi-supervised learning task, this strategy avoids the source domination
 thus facilitates more  effective domain adaptation.

\subsubsection{Metric Learning}\label{sec:metric_learning_v2_new}
Metric Learning within the context of semi-supervised joint training is distinctly innovative. In this scenario, \cite{semi4} adopts a metric learning strategy characterized by a disentangled paradigm. This approach separates style and content into distinct embedding spaces. Such separation facilitates independent contrastive learning for each aspect, allowing the model to adapt more effectively to variations in data distributions.

\subsection{Unsupervised Joint Training}\label{sec:unsupervised_joint_training}
Unsupervised Joint Training, commonly known as Unsupervised Domain Adaptation (UDA), is an advanced machine learning framework that facilitates the transfer of knowledge from a richly-labeled source domain to a completely unlabeled target domain. The central challenge of UDA lies in the absence of labels in the target domain, necessitating techniques that can align the underlying data distributions of both domains to enable accurate predictions on the target dataset. Key strategies include domain invariant feature extraction and distribution alignment.

\subsubsection{Cross-Modal Translation}\label{sec:cross_model_translation_2}
Cross-modal translation, employing techniques such as Generative Adversarial Networks (GANs) and frequency-based methods, is pivotal in transforming how we address domain differences by converting data from the source domain \( S \) to closely resemble the target domain \( T \). This transformation is formalized as:
\begin{equation}
\mathcal{\tilde{X}} = G(\mathcal{X}_s; \theta_G) \quad \text{where} \quad G: \mathcal{D}_s \rightarrow \mathcal{D}_t
\end{equation}
Here, \( G \) represents a generative model that minimizes domain discrepancies to align source domain \( \mathcal{D}_s \) with target domain \( \mathcal{D}_t \). The adaptation's effectiveness hinges on reducing the domain discrepancy metric \( d \), which measures differences between the adapted \( \mathcal{\tilde{X}} \) and target \( \mathcal{X}_t \):
\begin{equation}
\min_{\theta_G} d(\mathcal{\tilde{X}}, \mathcal{X}_t) = \min_{\theta_G} d(G(\mathcal{X}_s; \theta_G), \mathcal{X}_t)
\end{equation}
Once the data \( \mathcal{\tilde{X}} \) closely resembles \( \mathcal{X}_t \), it can be used to train models with source domain labels \( \mathcal{Y}_s \), significantly enhancing the model's ability to generalize across domains and mitigate image scarcity in specialized fields.

The application of GANs has evolved significantly, beginning with methods like the pixel-to-pixel (pix2pix) translation and advancing to more complex implementations. For instance, \cite{li2021restoration} utilizes an end-to-end unsupervised method for enhancing contrast in cataract fundus image based on pix2pix framework. Then, MADGAN~\cite{han2021madgan}  breaks the constraints of paired images, contributing to anomaly detection in complex brain structures. SASAN~\cite{tomar2021self} takes a further step that incorporating self-attention modules in its GANs, enhancing focus on specific anatomical details during image translation.
Subsequently, the utilization of CycleGANs~\cite{zhu2017unpaired} marks another significant advancement, enabling unpaired image translations for cross-domain chest X-ray disease recognition~\cite{tang2019tuna,sanchez2022cx} and hip joint bone segmentation~\cite{zeng2021semantic}. This translation process is further refined in~\cite{zou2020unsupervised} with dual-scheme (source-target/target-source)  fusion and~\cite{dong2020can} with attention mechanism. Integrating disentangled representations into GAN frameworks, as seen in~\cite{chen2021iosuda, peng2022unsupervised, wang2022cycmis, yang2019unsupervised, jiang2020unified, yao2022novel,cai2025style}, significantly advances domain adaptation by separating content from style, enhancing adaptation  efficiency. More recently, \cite{feng2025mamba} leverages a Mamba-based architecture to balance global context modeling with computational efficiency, further optimizing modality translation.

Complementing the adversarial nature of GANs, frequency-based methods~\cite{zhang2022cross, hu2022frequency} introduce a novel  perspective. They assume that the style information is stored in low frequency components and high frequency components represents more structural information, and thus translate the images  by replacing the low frequency components. Finally, techniques like singular value decomposition for noise adaptation in retinal OCT images~\cite{koch2022noise} highlight the innovation and adaptability in this field, which is  tailored to specific imaging modalities or diagnostic requirements.

\subsubsection{Pseudo-labeling}\label{sec:plabel_v2}
In UDA, as all target data labels are unknown, it becomes more challenging to make accurate pseudo-label predictions using traditional techniques. Research has since advanced the pseudo-labeling concept by integrating pseudo-labeling and adversarial learning to enhance the process~\cite{p1xie2022unsupervised}. Subsequent studies have built on this foundation, each offering unique improvements to address issues such as noisy labels~\cite{p4du2021constraint} and enhancing label reliability through methods like iterative self-training~\cite{p2shin2022cosmos}, contrastive learning~\cite{p6liu2022margin}, data augmentation~\cite{wu2024fpl+} and entropy constraints~\cite{p10feng2023unsupervised}. The specialized applications of pseudo-labeling are further explored in studies~\cite{p3li2022domain,p7cho2022effective,p9liu2021generative}. For example, \cite{p3li2022domain} focuses on nuclei instance segmentation and classification, utilizing pseudo-labels derived from prototype features. \cite{p7cho2022effective} breaks new ground in cell detection with a pseudo-cell-position heatmaps. \cite{p9liu2021generative} innovates by incorporating pseudo-labeling into tagged-to-cine MRI synthesis task, employing a Bayesian uncertainty mask for selective pseudo-label generation.

\subsubsection{Adversarial Learning}\label{sec:adlearn_v2}
Adversarial Learning is widely used for the implicit alignment between domains at feature or/and pixel level due to the absence of target labels.
At the feature level, techniques such as the plug-and-play adversarial domain adaptation network (PnP-AdaNet)~\cite{dou2019pnp} aligns features across different scales for segmentation tasks.  Similarly, \cite{jiang2022disentangled} aligns extracted contents for cross-modality segmentation. Other studies focus on prediction space alignment at the pixel level for various medical imaging tasks~\cite{liu2022ecsd,lei2021unsupervised,yan2019edge, shen2020unsupervised}.  Integrated approaches that apply adversarial training at both feature and output levels are explored in studies like~\cite{chen2019synergistic,chen2020unsupervised,shin2021unsupervised,liu2022bidirectional,su2023mind,calisto2022c,bian2022dda}. Innovations in this field also include enhanced discriminators and local discriminators that focus on specific region alignment, introducing spatial-aware and class-specific attentions to refine the adversarial loss and improve model adaptability across domains~\cite{pham2021unsupervised,yoo2021transferring,liu2021automated,chen2022dual}.

\subsubsection{Self-ensembling}\label{sec:selfensemble_2}
Initial studies by~\cite{1perone2019unsupervised} applied self-ensembling to gray matter MRI segmentation. Subsequent applications include breast MRI segmentation~\cite{3kuang2023mscda} and pose estimation in operating rooms~\cite{4srivastav2022unsupervised}. More advanced techniques combine adversarial training and self-ensembling for addressing domain shifts in cross-institutional gliomas studies~\cite{2shanis2019intramodality}, optic disc and cup segmentation~\cite{6liu2019cfea} as well as cardiac substructure segmentation~\cite{7zhao2022uda,cai2024symmetric}. 
Other significant developments include MT-UDA~\cite{5zhao2021mt}, which introduces a multi-teacher framework, and~\cite{8liu2023reducing,liu2023structure} further integrates frequency and spatial domain through multi-teacher distillation. Moreover, \cite{9hong2022unsupervised} explores a ``student-to-partner'' paradigm during various training stages.

\subsubsection{Graph-based Methods.}\label{sec:graph_based_1}
Graph-based methods are increasingly utilized in cross-domain medical image analysis due to their capability to capture complex spatial structures and relationships. This approach models image elements -- ranging from individual pixels to entire regions -- as nodes in a graph, with edges formed based on criteria like spatial proximity and similarity in intensity or texture.
The core of this method involves a graph  \( G = (V, E) \), with \( V \) representing the vertices and \( E \) the edges, which are weighted according to the mentioned criteria.  This setup facilitates the use of graph convolutional networks (GCNs)~\cite{kipf2016semi,velivckovic2017graph,guo2024gnn}, which leverage the graph structure for learning, described mathematically as:
\begin{equation}
H^{(l+1)} = \sigma \left( D^{-\frac{1}{2}} A D^{-\frac{1}{2}} H^{(l)} W^{(l)} \right)
\end{equation}
where \( H^{(l)} \) represents the features at each node in layer \( l \), \( A \), \( D \) is adjacency and degree matrix, \( W^{(l)} \) is the weight matrix for layer \( l \), \( \sigma \) denotes the activation function.  This process effectively leverages node features and graph topology for a comprehensive analysis.
Applications of this method include feature disentanglement~\cite{guo2022learning} for domain-invariant learning~\cite{graph2} with GCN, graph Laplacian decomposition for brain imaging alignment across domains~\cite{graph1}, attention-guided GCN for identifying major depressive disorder~\cite{graph4}, and a class-aware GCN classifier with domain-specific features for predicting lymph node metastasis in gastric cancer~\cite{graph3}.
Other notable implementations like~\cite{graph5}  extends beyond traditional methods by incorporating an online sub-graph scheme, 
\cite{graph6} employs GCNs with a meta-learning strategy targeting at   small-sized pancreatic cancer features. 
Studies like~\cite{graph7,graph8} focus on enhancing feature alignment and understanding inter-category relationships using graph-based techniques.

\subsubsection{Statistical Discrepancies Minimization}\label{sec:sta_dis_mini}
Quantifying and subsequently minimizing the statistical discrepancies between source and target domain feature spaces serves as a key approach. This paradigm, rooted in the hypothesis that reducing such discrepancies aids in model adaptation, predominantly employs measures like Kullback-Leibler (KL) divergence, Maximum Mean Discrepancy (MMD), and Correlation Alignment (CORAL). For example, the  adaptation with MMD~\cite{long2016unsupervised} between source and target domains can be mathematically formulated as:

\begin{equation}
\begin{split}
\mathcal{L}(D_s, D_t) & =  \frac{1}{n_s^2} \sum_{i,j=1}^{n_s} k(z_s^i, z_s^j) + \frac{1}{n_t^2} \sum_{i,j=1}^{n_t} k(z_t^i, z_t^j) \\
& - \frac{2}{n_s n_t} \sum_{i=1}^{n_s} \sum_{j=1}^{n_t} k(z_s^i, z_t^j)
\end{split}
\end{equation}

\noindent where \( k(z, z') = \exp\left(-\frac{\| \text{vec}(z) - \text{vec}(z') \|^2}{2b} \right) \) is the Gaussian kernel function defined on the vectorization of tensors \( z \) and \( z' \) with bandwidth parameter \( b \), $z_s, z_t$ are the multi-layer fused feature of the source and target domains.
This distance assess how similar or dissimilar the feature representations of the two domains are, and the goal is to adjust the feature representations such that the distance between the empirical distributions (as represented by the kernel functions) is minimized, thereby enabling adaptation.

In the medical adaptation field, \cite{gao2022unsupervised} explores the utility of MMD for domain adaptation in breast and thyroid lesions in ultrasound images.~\cite{hu2022domain} leverages KL divergence to synchronize  the prior distribution of the synthesized and the real target distribution. \cite{lu2022unsupervised} estimates the mutual information with KL divergence between the reconstruction output and segmentation result, so as to benefit each other. \cite{liu2022ordinal} enforces recursively conditional Gaussian (RCG) as the joint distribution prior, inheriting the closed form of the KL divergence term in the variational objective to make large-sacle tasks computationally tractable. 
\cite{xu2022dynamic}  uses MultiKernel Maximum Mean Discrepancy (MK-MMD) in aligning feature distributions in breast ultrasound images. Beyond these widely adopted metrics, novel metrics have been developed to suit specific medical tasks. \cite{wu2020cf} proposes the Characteristic Function (CF) Distance, transforming feature distributions to frequency domain for discrepancy calculations. \cite{sager2022unsupervised} introduces Domain Sanity Loss, focusing on anatomical features like centroid distance and plausibility in vertebrae prediction. \cite{zheng2024dual} introduces Inter-channel similarity Feature Alignment (IFA) loss, which aligns cross-channel distributions using structural priors and anatomical volume fractions.

\section{Federated Learning}\label{sec:Federated Learning}

\begin{figure}
  \centering
  \includegraphics[width=0.45\textwidth]{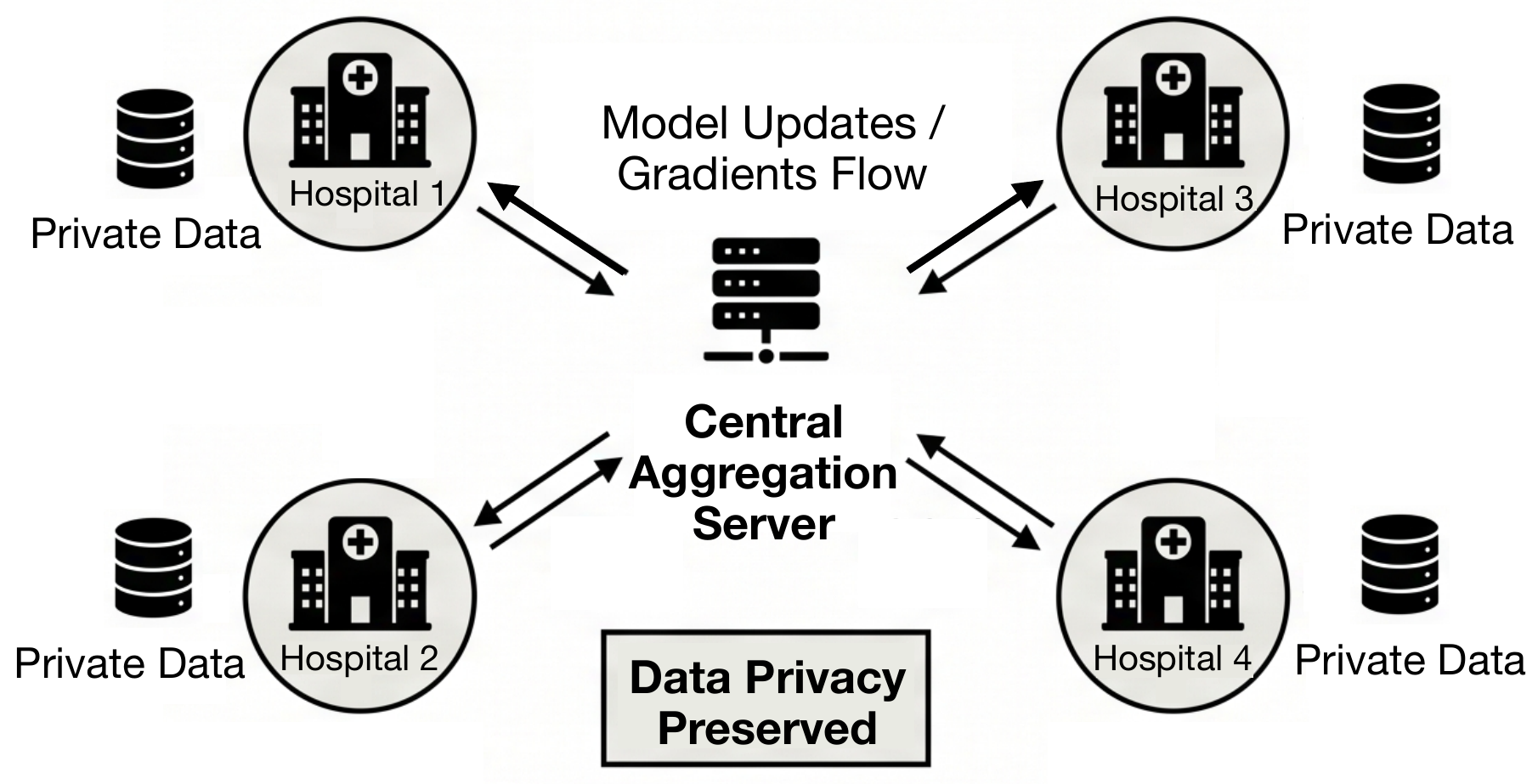}
  \caption{Federated Learning in MedIA. It facilitates collaborative model training across healthcare institutions while preserving data privacy. By maintaining data decentralization, it effectively addresses data distribution shifts without centralizing sensitive medical information, ensuring privacy and enhancing model robustness.}
\label{fig:banner_fl}
\end{figure}

Federated Learning is a pivotal model training approach designed to handle data heterogeneity while preserving the privacy of each client. It is particularly valuable in MedIA for alleviating data distribution shifts, allowing for collaborative enhancements across multiple healthcare institutions without the need to centralize their data (see Fig.~\ref{fig:banner_fl}). This decentralized method ensures the privacy of patient information, making it a practical solution for scenarios where medical data cannot be openly shared. One prominent example of Federated Learning in practice is FedAvg~\cite{fedavg}, which forms the basis for many modern implementations. In this model, each participating institution trains a local model on its own data, thereby maintaining the confidentiality of sensitive information. These institutions then send their model updates -- commonly in the form of weights or gradients -- to a central server. The server aggregates these updates to enhance the global model, which is shared back with all participating institutions after a few iterations. The mathematical formulation is detailed as follows:
\begin{align}
&\text{Local Update: } \theta_k^{(t+1)} = \theta_k^{(t)} - \eta \nabla L_k(\theta_k^{(t)})\\
&\text{Global Update: } \theta^{(t+1)} = \sum_{k=1}^K \frac{n_k}{n} \theta_k^{(t)}
\end{align}

\noindent where $\theta_k$ and $\theta$ respectively represent the parameters of the local model for the $k$-th client and the global model. Each client \( k \) contributes \( n_k \) data points, which together total \( n \) data points across \( K \) clients. The learning rate is denoted by \( \eta \), and \( \nabla L_k(\theta_k^{(t)}) \) refers to the gradient of the loss \( L_k \) with respect to the local model parameters at the \( k \)-th client. 

Similar to the Joint Training category, Federated Learning methods can also be classified based on the degree of data labeling. However, unlike Joint Training, which primarily focuses on the label availability of target data, Federated Learning treats both source and target data as clients that play similar roles. Each client trains a local model that contributes to the quality of the expected global model. Therefore, in this section, our primary concern regarding label availability extends to all clients. Based on this, Federated Learning for MedIA under distribution shifts can be divided into Supervised and Semi-supervised Federated Learning. Besides, it is worth noting that, given that medical settings often feature well-characterized source datasets, the need for Unsupervised Federated Learning approaches~\cite{lubana2022orchestra,kim2023protofl} is generally minimal.

\subsection{Superivsed Federated Learning}
Early research in the medical field utilizing the FedAvg~\cite{fedavg} algorithm targeted a wide range of medical tasks, from Brain leison Segmentation~\cite{fedapp1,wagner2025feasibility} to the detection of COVID-19 lung abnormalities~\cite{fedapp2,fedapp6}, MRI Reconstruction~\cite{fedapp4}, Diabetic Retinopathy Classification of OCT Data~\cite{fedapp7}, and Breast Cancer Histopathological Image Classification~\cite{fedapp8}. These initial applications laid the foundation for using Federated Learning to process sensitive medical data across distributed datasets while maintaining privacy. Further studies explored the influence of factors like the number of healthcare providers, dataset size, non-IID sampling, communication strategies, and architecture types on FL performance in medical contexts~\cite{fedapp5,fedapp10}. As the field progressed, benchmarks were established to assess the effectiveness of various FL algorithms in managing data heterogeneity~\cite{fedapp11} across diverse medical datasets. Moving beyond the basic FedAvg paradigm, current advancements have focused on addressing issues like Data Heterogeneity and Client Drift~\cite{fedprox,fedrep}, which arise from non-IID data distributions among clients. These shifts can significantly affect model performance, prompting researchers to develop strategies for the accuracy of FL models in healthcare settings.

\subsubsection{Data Augmentation}\label{sec:data_aug_1}
In Federated Learning for MedIA under distribution shifts, cross-client data augmentation plays a crucial role in managing the inherent diversity and imbalance of data across different clients. This strategy is designed to enhance the uniformity of feature representations across participating clients, thus improving the overall robustness and accuracy of the federated model. Techniques such as Fourier transform-based methods~\cite{feddata9} are particularly effective, as they allow for the sharing and interpolation of frequency domain information among clients, promoting a more consistent feature representation across varied datasets. Specifically, HarmoFL~\cite{feddata25} leverages frequency information to unify amplitude components across clients, which aids in maintaining consistent low-level visual features. Other systematic augmentation techniques~\cite{feddata27, feddata29, feddata30, feddata16, xie2024pflfe} explore various augmentation strategies to combat data diversity and imbalance. These techniques vary in their approaches but collectively contribute to a more equitable and effective training process, enhancing the ability of federated models to generalize across diverse environments and data conditions.

\subsubsection{Novel Architecture Design}\label{sec:novel_arch_design_v1}
Some strategies specifically address distribution shifts by developing tailored model architectures. For instance, SU-Net~\cite{feddata34} enhances standard U-Net with inception modules and dense blocks to manage multi-scale challenges effectively. Similarly, FedDAvT~\cite{feddata15} leverages Transformer architecture to facilitate domain adaptation for Alzheimer’s disease diagnostics. 
Adversarial and generative networks are introduced to refine federated learning, focusing on aligning or adapting feature spaces across different clients~\cite{feddata21, feddata17}. 

\subsubsection{Novel Training Strategies}\label{sec:novel_train_stra_v2}
Novel training strategies are being explored to enhance the efficacy and adaptability of models under distribution shifts. For example, FedSM~\cite{feddata6} optimizes model selection during inference, while FedCross~\cite{feddata7} utilizes sequential training to bypass model aggregation. Building on these concepts, FedVCK~\cite{yan2025fedvck} employs dataset condensation to synthesize logit prototypes, and MSAFed~\cite{jin2025msafed} leverages multi-level contrastive learning with prototype-aware aggregation to handle heterogeneous distributions. Further advancing the structural flexibility, FedCGP~\cite{niu2025fedcgp} clusters similar clients to separate shared knowledge from client-specific features, while PAFedMIS~\cite{li2025pafedmis} decouples encoder-decoder structures to facilitate asynchronous personalized aggregation. To further decentralize the process, GCML~\cite{chen2025decentralized} replaces the central server with gossip-based model exchange and introduces contrastive mutual learning to reduce negative transfer. Beyond architectural shifts, FEAL~\cite{chen2024think} incorporates evidential uncertainty into active learning to ensure reliable data selection. These innovative methods, alongside techniques like Mixture of Experts and Split Learning~\cite{feddata8, feddata23}, collectively contribute to more resilient model training and deployment.

\subsubsection{Metric Learning.}\label{sec:metric_learning_v2}
Several methods utilize metric learning to enhance consistency between different clients in federated settings. For instance, FedIIC~\cite{feddata4} implements two-level contrastive learning to optimize both intra- and inter-client feature consistency, ensuring uniformity in the learned representations. FedCL~\cite{feddata14} focuses on reducing the feature distance between successive local and global  models, which helps stabilize the training process. Similarly, FedDP~\cite{feddata3} improves model uniformity across clients by penalizing inconsistencies during the learning phase. Additionally, LC-Fed~\cite{feddata5} employs contrastive site embedding and makes prediction-level adjustments to enhance personalization.

\subsubsection{Aggregation Weight Calibration}\label{sec:agg_weig_calb}
In Federated Learning for MedIA, aggregation weight calibration is a sophisticated optimization strategy that refines how global model updates are weighted, taking into account more than data volume. This method involves adjusting the influence of each client's local update on the global model by considering factors such as the stage of training, client performance, and similarity between client models and the global model. For example, \cite{feddata10,feddata26} highlight strategies where weights are calibrated based on the training progress and the performance metrics of clients. Additionally, the similarity-based approach~\cite{feddata11, feddata12, feddata28, feddata1, jiang2023fair, chen2024fedevi} assesses how closely aligned each client’s data distribution or model parameters are with the global model. This alignment influences their weights during aggregation, promoting updates that are more representative of the overall data characteristics. Moreover, FedAWA~\cite{feddata24} introduces an innovative twist by employing reinforcement learning to dynamically adjust client weights. This system continually learns and updates based on data distribution and feedback from client performance, optimizing the aggregation process to ensure the global model remains robust and accurate across varying conditions.

\subsubsection{Parameter Calibration}\label{sec:param_clab}
Parameter calibration also plays a crucial role, specifically for addressing the conflict between the local and global models. It involves strategically adjusting model parameters to ensure that the collective learning benefits all participating clients. Efforts include rescaling local parameters~\cite{feddata13} and mixing local and global gradients~\cite{feddata2} to enhance model convergence and stability. \cite{feddata33} proposes a Deputy-Enhanced Transfer strategy at the client site. It firstly leverages a deputy model to  receive aggregated parameters from the server, and then smoothly transfers the global knowledge to the personalized local model.
Some other strategies emphasize fairness, such as those aiming to equalize training loss by adjusting the model parameters such that all hospitals have a similar training loss~\cite{feddata31}. This approach ensures that no single client's data disproportionately influences the model, thus maintaining uniformity in model performance regardless of the data source.

\subsection{Semi-supervised Federated Learning}
In the diverse landscape of Federated Learning, Semi-Supervised Federated Learning (SSFL) emerges as a pivotal area of exploration, particularly suited to complex environments like healthcare, where only a subset of clients possess fully labeled data, while a significant portion operates with unlabeled datasets. 
By incorporating techniques from semi-supervised learning, SSFL effectively utilizes sparse labels to extrapolate knowledge and enhance learning from the extensive unlabeled data available. 
This approach not only broadens the applicability of Federated Learning in the medical field but also adeptly addresses the latent data heterogeneity challenges that emerge when the lack of clear labels obscures underlying data variations.

\subsubsection{Pseudo-labeling}\label{sec:peseudo_labeling_v3}
Several innovative approaches have been developed to enhance the utility of pseudo-labeling in federated settings. \cite{fedspe4} introduces a novel method that integrates prototype-based pseudo-labeling with contrastive learning, a technique also employed by~\cite{fedspe6}. Additionally, \cite{fedspe7} enhances pseudo-label generation by incorporating a self-supervised rotation loss, which provides consistent regularization across unlabeled datasets. Further, \cite{fedspe2} improves the connection between labeled and unlabeled data by aligning disease relationships across clients, effectively compensating for the lack of task-specific knowledge in unlabeled clients and enhancing the extraction of discriminative information from unlabeled samples.

\subsubsection{Transformer-based Architecture}\label{sec:transformer_arch_v2}
Transformer offers a robust framework for leveraging both labeled and unlabeled data within a single client. For example, \cite{fedspe12}~exemplifies a specialized approach where a self-supervised learning framework is implemented using Transformer architectures. This method starts with masked image modeling, a self-supervised task that trains the model to predict the portions of images that are intentionally obscured. This phase harnesses the abundant unlabeled data, allowing the model to learn rich, generalized features without requiring too many explicit labels.

\subsubsection{Novel Training Strategies}\label{sec:novel_training_stratg_v3}
\cite{pan2025dual} introduces a dual-calibrated co-training approach to address the dual challenges of data heterogeneity and semi-supervised inadaptability in medical image segmentation. By constructing a rule relationship graph that considers model similarity, dataset size, and quality, the system aggregates a unique personalized model for each client to effectively adapt to unique local feature distributions. 

\subsubsection{Advanced Optimization Strategies}\label{sec:advance_opt_stratg}

SSFL also address the dual challenges of data scarcity and distribution heterogeneity.
One innovative approach is the Federated Drift Mitigation (FedDM) framework~\cite{fedspe5}, which achieves robust gradient aggregation by resolving conflicts between gradients at different network layers, as guided by the historical gradients of the global model. Another strategic implementation is FedCy~\cite{fedspe14}, designed for surgical phase recognition. This method integrates dual training objectives: it applies consistency learning to exploit the temporal and spatial consistencies in the unlabeled data, alongside contrastive learning techniques to enrich the learning from sparsely labeled data.

\section{Fine-tuning}\label{sec:Fine-tuning}
\begin{figure}
  \centering
  \includegraphics[width=0.45\textwidth]{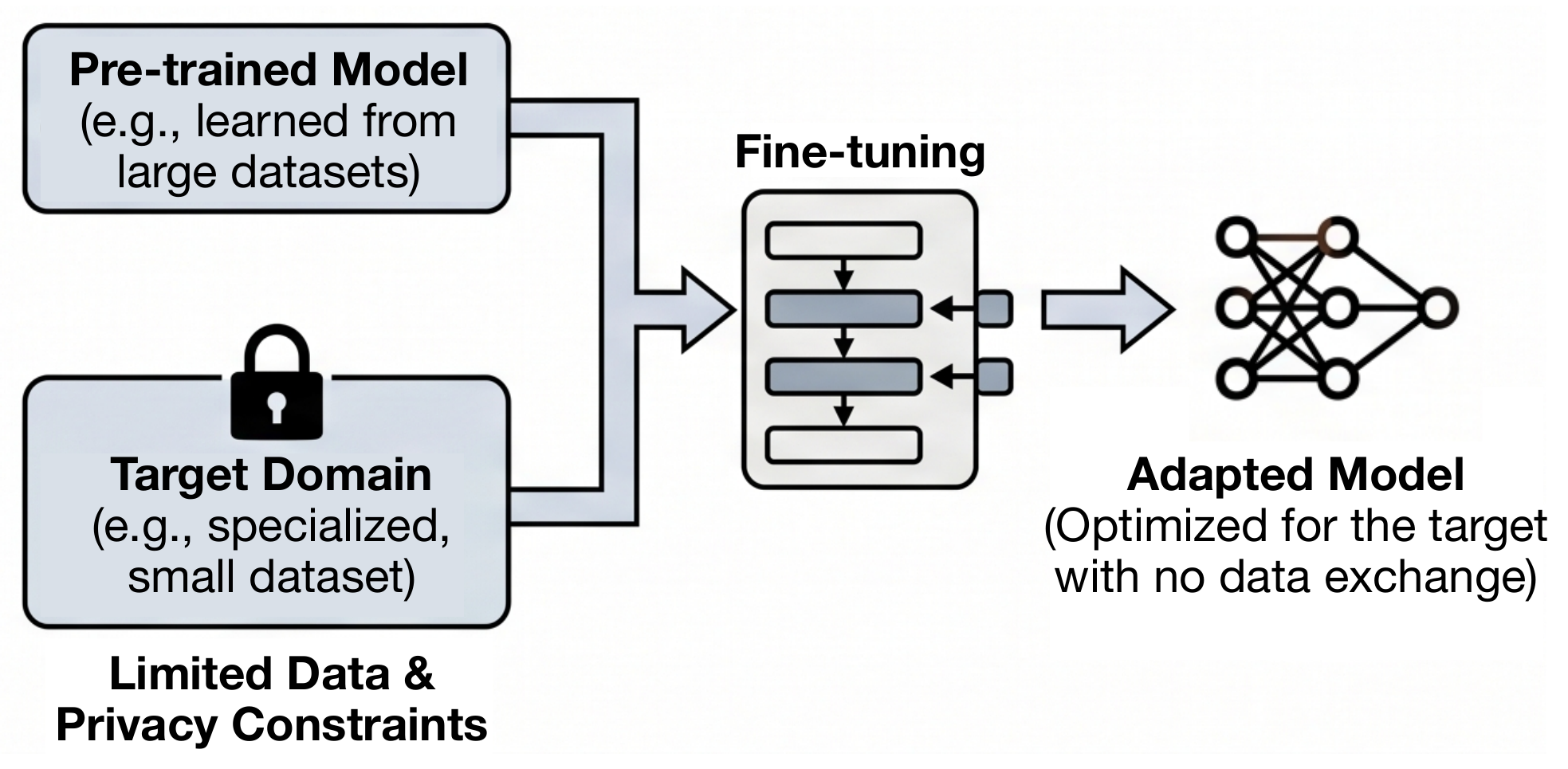}
  \caption{Fine-tuning in MedIA: It adapts pre-trained models to specialized datasets, particularly when privacy concerns limit data sharing and direct collaborations are impractical. This approach enables healthcare institutions to leverage the valuable prior knowledge with minimal data exchange.}
\label{fig:banner_finetuning}
\end{figure}

Fine-tuning plays a vital role in enhancing the adaptability and performance of pre-trained models across a wide range of applications. This process involves adjusting a model that has been pre-trained on a large, generic (source) dataset to perform effectively on a different, often smaller and more specialized (target) dataset. In medical scenarios, Fine-tuning proves particularly effective when privacy concerns preclude open data sharing, and synchronous collaborations among different healthcare institutions are impractical or excessively costly. This strategy enables medical institutions to leverage pre-existing models and adapt them with minimal data exchange, effectively addressing privacy and collaboration constraints in MedIA, as illustrated in Fig.~\ref{fig:banner_finetuning}.
Based on the availability of labeled data on the target domain, fine-tuning methods are classified into supervised and unsupervised approaches. As we move from supervised to unsupervised settings, the complexity increases but so does the significance of the application, offering broader adaptability to real-world challenges where labeled data are limited.

\subsection{Supervised Fine-tuning}
\label{sec:supervised Fine-tuning}

Supervised Fine-tuning stands out as a potent method for enhancing diagnostic accuracy in MedIA. This technique primarily involves applying specific pre-trained networks, such as VGG~\cite{vgg} and AlexNet~\cite{alexnet}, initially trained on general images like ImageNet~\cite{deng2009imagenet}, to more specialized medical imaging tasks. Research exemplified by studies~\cite{finetune2,finetune3,finetune4,finetune6} demonstrates how these models transition to applications in medical imaging, including tumor classification and chest X-ray analysis, leveraging their capability to generalize features across diverse visual domains for precise medical diagnostics. Fine-tuning these networks often requires minimal adaptation design, making it a straightforward approach to boost performance in medical tasks. Notable successes also include adapting networks for Alzheimer's diagnosis~\cite{finetune1} and employing the Med3D network for detailed lung segmentation and nodule classification~\cite{finetune5}.

\subsubsection{Novel Strategies and Structures}\label{sec:Novel Strategies and Structures}
Beyond simply evaluating on different pre-trained network architectures, some research have focused on novel strategies and structures for rapid and accurate domain adaptation, while preserving existing knowledge. \cite{finetune11} introduces ContextNets, a memory-augmented network for seamless domain adaptation in semantic segmentation without the need for extensive retraining. In contrast, \cite{finetune10} employs Elastic Weight Consolidation to maintain performance by encoding information from previous tasks, without extra data storage. Furthermore, \cite{finetune12} optimizes batch normalization to swiftly adjust to new domains while maintaining shared convolutional layers across all domains.

\subsection{Unsupervised Fine-tuning}
\label{sec:Source-Free DA}
Unsupervised Fine-tuning in MedIA is an innovative response to the constraints of traditional supervised fine-tuning that rely heavily on labeled target datasets which are often unavailable in healthcare scenarios. This approach, crucial in healthcare where rapid adaptation is required to varying patient data, is characterized by two primary branches: Source-Free Domain Adaptation (SFDA)~\cite{sfda_kdd} and Test-time Adaptation (TTA)~\cite{wang2021tent}. Both are designed to adapt models dynamically to new and changing conditions without the need for source data at the time of inference, thus directly addressing the challenges of data privacy. SFDA achieves this by transferring knowledge learned during training and applying it to new test samples through adaptive modules or auxiliary self-supervised tasks, such as rotation prediction. This allows the model to train on the target distribution for multiple epochs before making predictions, providing a proactive adaptation approach. On the other hand, TTA takes on a more challenging task by requiring the model to adapt in real-time to a continuous stream of test data, making no prior adjustments during the training phase. This method is model-agnostic and focuses on immediate, on-the-fly adjustments to effectively process and respond to incoming data. Both strategies share the common goal of enabling efficient model adaptation in unsupervised settings, ensuring that medical diagnostics remain robust and accurate even when faced with data that significantly deviates from previously seen examples.

\subsubsection{Pseudo-labeling}\label{sec:pesu-label-v3}

An intuitive solution for SFDA/TTA is to generate pseudo labels~\cite{TTA7, TTA8, TTA15, TTA12} to convert the target adaptation into a supervised task. However, these labels often contain noise, necessitating refinement through confidence and geometric constraints. Early methods include using dual-classifiers to enhance label confidence~\cite{SFDA9}, applying denoising via uncertainty and prototype distance estimation~\cite{SFDA5}, and employing shape compactness metrics for reweighting~\cite{SFDA7}. Recent efforts further expand the scope of refinement; for instance, \cite{wu2025a3} incorporates high-confidence sample selection and memorization for label fusion, while \cite{SFDA12} integrates image quality and irregular structure detection to select optimal training labels.
Moving towards predictive stability, \cite{TTA18} utilizes the greatest union mask of multiple predictions as proxy labels. This concept of stability is further explored by \cite{yang2023tgma} through consistency checks under masked perturbations, and by \cite{zhou2025tegda} via dropout-based agreement and feature fusion with a high-quality sample bank. Finally, \cite{TTA12} selects low-entropy pixels and applies contrastive learning to tighten target feature distributions. Collectively, these strategies mitigate the impact of domain discrepancies by iteratively purifying the training signal.

\subsubsection{Image Generation}\label{sec:image_cap}
Image generation techniques facilitate the adaptation of models to new domains by enriching the dataset with varied and representative examples. 
For example, \cite{SFDA10} utilizes basic image augmentation combined with causal interventions to generate diverse datasets that ensure consistent predictions and the elimination of confounding factors. Similarly, \cite{SFDA11} employs patch-wise processing augmented with a Transformer structure to enhance data variability effectively, while~\cite{TTA2} proposes the first learnable test-time augmentation policy that dynamically selects most effective augmentation techniques. This adaptability allows for optimal model performance even under varying operational conditions. Moreover, some strategies focus on transforming the style of data between the source and target domains to better align the characteristics of the target data with the learned source domain model. 
\cite{SFDA4} applies autoencoders to make test images resemble source data, while \cite{SFDA6} uses class-conditional GANs to synthesize target-style images. A common approach in this category is Fourier-based style mining; \cite{SFDA7} and \cite{huang2023fourier} both leverage Fourier amplitude swapping to transform images into source-like styles. More recent strategies utilize domain-aware prompts for input-level modification. While \cite{TTA14} learns a general prompt to match source styles, VPTTA~\cite{chen2024each} and DDFP~\cite{yin2025ddfp} refine this by learning per-image prompts that modify low-frequency components. Building on this, EDCP~\cite{liu2025efficient} incorporates a lightweight convolution-plus-offset module as a prompt, updated in real-time via structural and layer-wise alignment losses. These innovative methods ensure smoother adaptation by directly bridging the stylistic gap at the input level.

\subsubsection{Batch Normalization}~\label{sec:batch_normalization}
Batch Normalization has been widely explored in adaptation tasks as normalization statistics are associated with the domain  distribution. They can be directly obtained through pre-trained model and taken as the source information. 
Given a mini-batch $\mathcal{B} = \{x_n\}_{n=1}^N$ where $x_n \in \mathbb{R}^{F}$ is a feature vector (with $F$ denoting the number of feature channels and $N$ the batch size), BN normalizes each feature dimension $f$ as follows: 
\begin{equation}
\hat{x}_{n,f} = \frac{x_{n,f}-\mu_{\mathcal{B}, f}}{\sqrt{\sigma_{\mathcal{B}, f}+\epsilon }} \cdot \gamma_f  + \beta_f
\label{BN}
\end{equation}
where $\mu_{\mathcal{B},f}$ and $\sigma_{\mathcal{B}, f}$ are the running mean and variance for the $f$-th feature of mini-batch $\mathcal{B}$, respectively. The parameters $\gamma_f$ and $\beta_f$ are the learned scale and shift factors for affine transformation, with $\epsilon$ being a small-offset to avoid division by zero. \cite{TTA1} proposes an exponential decay scheme for the normalization statistics in adaptation stage to gradually learn the target domain-specific mean and variance.
\cite{TTA14} aligns the source and target normalization statistic discrepancy for learning a prompt to make the target inputs be treated as the source.
More recently, \cite{TTA5} explores domain-specific and shareable batch normalization statistics for adaptive BN-based adaptation, while~\cite{su2024unraveling} proposes to incorporate the concept of class diversity to address more realistic mini-batch problem.

\subsubsection{Novel Strategies and Structures}~\label{sec:Novel_Strategies_and_Structures_v2}
The field has seen several structural innovations aimed at overcoming specific  adaptation challenges. {For structure innovations,} \cite{TTA6} introduces an auxiliary rotation classifier to improve adaptation via self-training. Similarly, \cite{SFDA3} utilizes multiple diverse classifiers to address test label distribution shifts, and~\cite{TTA17} employs decoder duplication during the adaptation stage to ensemble diverse target inputs. \cite{zhu2025improving} employs a mean-teacher framework in which the teacher generates uncertainty-aware pseudo-labels to guide the student’s adaptation to unlabeled target data.
Y-shaped architectures with dual decoders are used for enhanced denoising and segmentation~\cite{SFDA15, SFDA1}. ~\cite{TTA4} further develops a supplementary network to  adaptively combined with the main outputs during inference.  
Recent frameworks further refine at the parameter and prompt levels. \cite{zhang2025enhancing} optimizes parameter-level transferability by stabilizing domain-invariant parameters while updating high-variance embeddings through whitening during self-training. Parallel to this, PASS~\cite{zhang2024pass} introduces a test-time prompting framework that jointly models style and semantic shape shifts using an input decorator and a cross-attention prompt bank.

\subsubsection{Entropy Minimization}\label{sec:entropy_min}
Entropy minimization is  widely-used to handle unlabeled data. 
Mathematically, the entropy of a prediction can be expressed as follows:
\begin{equation}
H(p) = -\sum_{i=1}^C p_i \log p_i
\end{equation}
where \( p \) represents the predicted probability distribution over \( C \) classes, and \( p_i \) is the probability of the \( i \)-th class predicted by the model. The goal is to minimize this entropy \( H(p) \) across the dataset, thereby encouraging the model to produce more decisive outputs.
This approach is first introduced by Tent~\cite{wang2021tent} into general TTA tasks, which proposes minimizing the mean entropy over the test batch to update the affine parameters of the batch normalization layers in the pre-trained model. This strategy has been adopted in many cases~\cite{SFDA8,TTA1,TTA4,TTA9,TTA10} in medical TTAs. 

\subsubsection{Distribution Discrepancies Minimization}\label{sec:distribution_discrep_mini}
Capturing the source distribution during training and aligning it with the target domain during adaptation constitutes a key approach to enhancing SFDA performance. \cite{stan2024unsupervised} models source decoder embeddings as a GMM after training and performs adaptation by aligning target embeddings to GMM samples via sliced Wasserstein distance.

\subsubsection{Dynamic Adjustment of Learning Rates}\label{sec:Dynamic Adjustment of Learning Rates}
Dynamic adjustment of learning rates based on distribution shifts helps models adapt more effectively during test-training stages. For example, \cite{TTA3} proposes that samples with larger distribution shift should result in larger update. It makes adjustment by calculating the divergence between the model outputs and its nearest neighbors in a memory bank. \cite{SFDA14} refines this strategy  by assessing category-wise discrepancies with an uncertainty estimation module.

\subsubsection{Anatomical Information.}\label{sec:anato_inform}
Leveraging the anatomical information as a prior for loss design offers a promising direction for enhancing the accuracy and reliability. For instance, \cite{SFDA2} utilizes a shape dictionary, integrating general semantic shapes extracted from source data. ~\cite{SFDA13} incorporates the shape information with the signed distance field which measures the distance between any pixel to the nearest object boundary and the relative  position.\cite{lv2025test} converts multi-source images into graph structures to learn a shared morphological node space, aligning target data via structural adaptation signals. \cite{TTA9,TTA10} leverages the class-ratio as a supervision which is estimated from anatomical knowledge available in the clinical literature. \cite{TTA13} uses anatomically-derived loss functions that penalizes unrealistic bone lengths and joint angles in 3D pose estimation. \cite{TTA16} proposes contour regularization loss for constraining the continuity and connectivity. 
\cite{SFDA16} expands the lesion click (i.e., the center of the nodule)  into an ellipsoid mask, and use it as the supervised information for test training.

\begin{figure}
  \centering
  \includegraphics[width=0.45\textwidth]{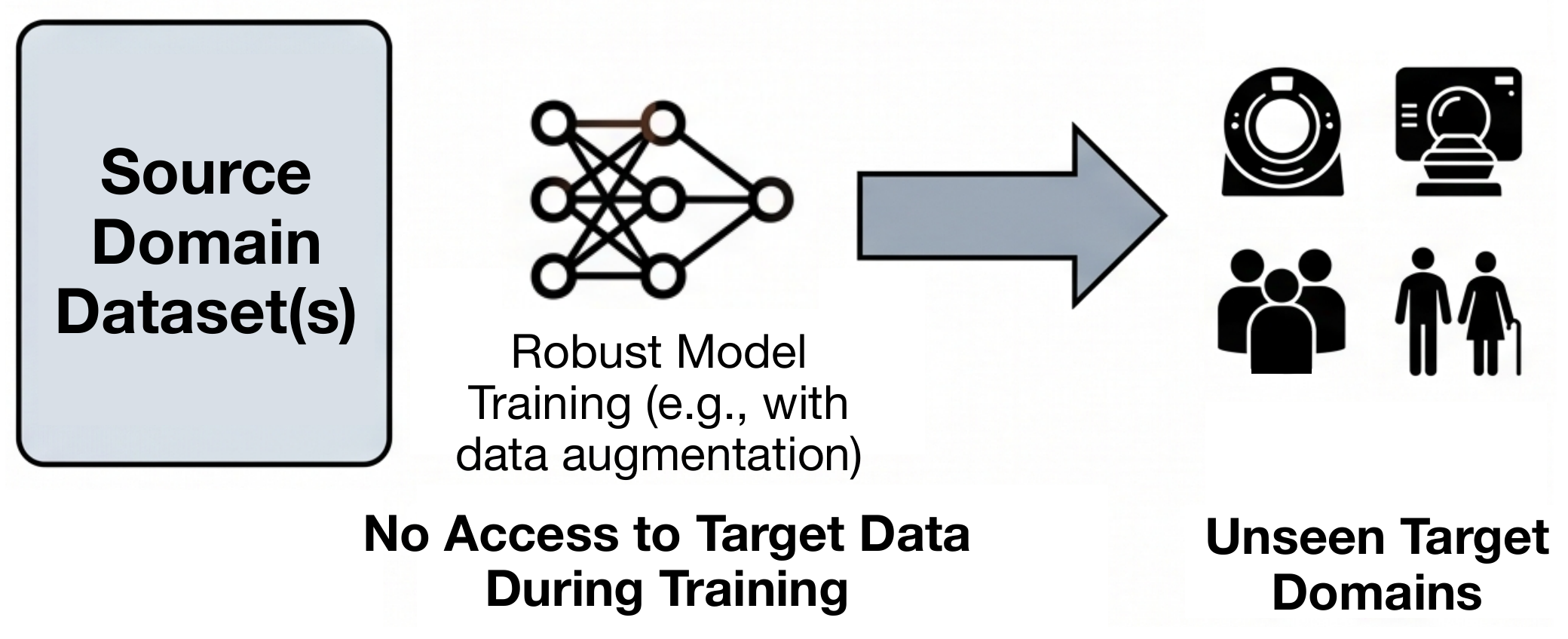}
  \caption{Domain Generalization in MedIA: It prepares models for unseen data by generalizing from source datasets without target data access. This proactive approach ensures robustness in unfamiliar environments. The shown flowchart is Single-source Domain Generalization.}
\label{fig:banner_dg}
\end{figure}

\section{Domain Generalization}\label{sec:Domain Generalization}
Domain Generalization (DG) is an advanced deep learning technique designed to prepare models for handling unseen, out-of-distribution data. This challenge is especially relevant in MedIA, where real-world operational constraints often make target datasets from new domains inaccessible or unknown. In such cases, DG methods become essential, enabling models to generalize from available source data to new environments without prior exposure to specific target data (see toy example in Fig.~\ref{fig:banner_dg}). This proactive approach ensures that medical models remain robust and accurate, ready to cope with potential unfamiliar environments. Furthermore, unlike other categories that are typically divided from simple to complex based on the label availability of target data, DG assumes that no target data is available. The complexity of tasks within this category primarily hinges on the nature of the source data. Specifically, DG techniques can be divided into two main types: Multi-source Domain Generalization (MDG) and Single-source Domain Generalization (SDG). MDG capitalizes on the diversity of multiple source datasets to extract and decouple domain-invariant and domain-specific features, thereby enhancing the model's generalizability using the domain-invariant component. Conversely, SDG, limited to a single source, faces greater challenges and often relies on additional data augmentation strategies to increase the model's generalization capabilities under more restrictive conditions.

\subsection{Multi-source Domain Generalization}\label{sec:Multi-source DG}

Multi-source Domain Generalization (MDG) operates under the premise that the unseen target domain shares the commonalities with the source dataset. The main challenge here is effectively extracting and balancing domain-invariant features -- which apply across all datasets -- and domain-specific features -- which are unique to each dataset. Techniques such as feature disentanglement and meta-learning are often employed to address these challenges, helping to enhance the model's ability to generalize while reducing the risk of overfitting to any single source domain.

\subsubsection{Meta-learning}\label{sec:meta_learning_v2}
Meta-learning~\cite{hospedales2021meta} is a powerful strategy for enhancing model generalization across unknown data distributions. This approach involves simulating domain shifts during training through ``episodes'', where data from multiple sources is split into meta-train $\mathcal{D}_{\text{train}}$ and meta-test $\mathcal{D}_{\text{test}}$ sets. This split mirrors real-world domain shifts, preparing the model for new domains or distributions. The model first learns from the meta-train set and is then tested on the meta-test set to evaluate its adaptability to new situations. Adjustments are made based on its performance to enhance its generalization capabilities.
\noindent
This process is formulated as:
\begin{equation}
\small
\phi^* = \text{\textit{MetaLearn}}(\mathcal{D}_{\text{train}}), 
\theta^* =  \text{\textit{Learn}}(\mathcal{D}_{\text{test}}; \phi^*)
\end{equation}
where $\phi^*$ denotes the meta-learned parameters, which are then used to learn the task-specific model parameters $\theta^*$ on the meta-test set. 
Following this framework, \cite{MDG4} introduces a shape-aware meta-learning scheme that incorporates anatomical integrity, \cite{MDG5} combines meta-learning with style-feature flow generation for confounding factors elimination, and~\cite{MDG3} uses style-transferred images as meta-tests, designing a new  boundary-oriented objective for meta-optimization considering the specific challenges in medical image segmentation.

\subsubsection{Shape-based Regularization}\label{sec:Shape-based Regularization}
Shape-based Regularization is a powerful tool, harnessing the continuous and coherent nature of anatomical structures and the domain-invariant characteristics of their contours. 
Except for combined with the meta-learning approaches~\cite{MDG4,MDG5,MDG3} for supervision in meta-test optimization, some methods directly use the anatomical knowledge as prior information during training.  For example, \cite{MDG2} integrates fixed Sobel kernels for contour enhancement and a convolutional autoencoder for learning anatomical priors, which inversely projects the mask and prediction to the feature space for further alignment.

\subsubsection{Latent Space Regularization}\label{sec:Latent Space Regularization}
Latent Space Regularization focuses on modeling inter-domain relationships and perform regularization in the latent feature space to promote generalization. Notably, \cite{MDG6} introduces a rank regularization term to constrain the complexity of  feature representations and restrict the latent features to follow a pre-defined prior distribution, while~\cite{MDG7} implements semantic feature regularization during the meta-test phase with dual losses that maintain global inter-class relationships and tighten intra-class features.

\subsection{Single-source Domain Generalization}\label{sec:Single-source DG}

Single-source Domain Generalization (SDG) presents a unique set of challenges as it relies on data from only one source to prepare models for unseen domains. This restriction is particularly pronounced in the medical field, where variability in data can be extreme and the stakes of accurate generalization are high. The primary challenge in SDG is the limited diversity, which can make models prone to biases and over-fitting, reducing their ability to perform well on novel, out-of-distribution medical data. To combat this, SDG strategies often incorporate robust data augmentation techniques -- such as synthetic image generation, geometric transformations, and intensity variations -- to artificially expand the dataset's diversity and simulate potential unseen scenarios. Additionally, regularization techniques and invariant feature learning are used to further enhance the model's generalization capabilities.

\subsubsection{Pixel-level Augmentation}\label{sec:Pixel-level Augmentation}
Pixel-level Augmentation techniques directly manipulate the pixel values. This method is primarily based on the premise that variations in imaging modalities, acquisition protocols, and hardware can induce significant discrepancies in image characteristics such as texture, intensity, and contrast. 
For example, \cite{SDG8} introduced BigAug, a deep stacked general transformation approach to systematically evaluate augmentation effects on model generalization. Specialized approaches within the medical field, such as the use of Bézier Curves by~\cite{SDG2} to address gray-scale discrepancies, and the simulation of MRI distortions by~\cite{MDG1}, focus on medical-specific image traits. \cite{SDG7} uses causal inference methods  to reflect acquisition shifts and~\cite{SDG6} explores category-level augmentation based on class-level representation invariance. \cite{SDG5} combines the augmentation strategies both in~\cite{SDG8} and~\cite{SDG6}. 
\cite{SDG3} expands the style space through adversarial training and finds the worst-case style composition to generate the samples. \cite{SDG4} further refines this strategy by  introducing randomness to the generated domain through a adaptive instance normalization block, so that the changes are limited to the textures.

\subsubsection{Feature-level Augmentation}\label{sec:Feature-level Augmentation}
Some augmentation strategies delves deeper into the model's internal workings, focusing on the manipulation of learned feature representations. 
For instance, \cite{SDG1} masks features channel-wisely and spatially to generate diverse challenging samples.

\begin{table*}[!t]
\centering
\caption{Details of cross-domain segmentation datasets used in this survey.}
\label{tab:data_info}
\renewcommand{\arraystretch}{1.6}
\setlength{\tabcolsep}{1pt}
\resizebox{\textwidth}{!}{
\begin{tabular}{c|c|c|c|l|c} 
\toprule
Dataset & \centering{Task} & Label(s) & Split & \multicolumn{1}{c|}{Domain(s)} & No. of Samples \\ \hline

\multirow{2}{*}{\parbox{3cm}{\centering Cardiac MRI-CT \\ \cite{CARDIACDATASET}}} & 
\multirow{2}{*}{
\parbox{4.5cm}{
\justifying
Unsupervised Domain Adaptation (UDA)
}
} & 
\multirow{2}{*}{\parbox{4.2cm}{\centering Ascending Aorta, Left Atrium Cavity, Left Ventricle Blood Cavity, Myocardium}} & 
Source & b-SSFP MRI & 20 \\
& & & Target & Computed Tomography (CT) & 20 \\ \hline

\multirow{2}{*}{\parbox{3cm}{\centering Abdominal CT-MRI \\ \cite{abdominalCT, abdominalMRI}}} & 
\multirow{2}{*}{
\parbox{4.5cm}{
\justifying
Source-Free Domain Adaptation (SFDA) \& Test-Time Adaptation (TTA)
}
} & 
\multirow{2}{*}{\parbox{3.5cm}{\centering Liver, L-kidney, \\ R-kidney, Spleen}} & 
Source & Computed Tomography (CT) & 30 \\
& & & Target & T2 SPIR MRI & 20 \\ \hline

\multirow{2}{*}{\parbox{3cm}{\centering Prostate Cross-center \\ \cite{MDG4, bloch2015nci, lemaitre2015computer, litjens2014evaluation}}} & 
\multirow{2}{*}{
\parbox{4.5cm}{
\justifying
Single-source Domain Generalization (SDG)
}
} & 
\multirow{2}{*}{Prostate} & 
1 Source & \multirow{2}{*}{Prostate MRI from 6 centers} & 30, 30, 19 \\
& & & 5 Targets & & 13, 12, 12 \\ \hline

\multirow{2}{*}{\parbox{3cm}{\centering Fundus Cross-center \\ \cite{almazroa2018retinal, fumero2011rim, orlando2020refuge, sivaswamy2015comprehensive}}} & 
\multirow{2}{*}{
\parbox{4.5cm}{
\justifying
Federated Learning (FL)
}
} & 
\multirow{2}{*}{Optic Disc, Cup} & 
\multirow{2}{*}{6 clients} & \multirow{2}{*}{Retinal fundus images from 6 centers} & 101, 159, 400 \\
& & & & & 400, 85, 195 \\

\bottomrule
\end{tabular}}
\label{tab: dataset_detail}
\end{table*}

\begin{figure*}[t]
  \centering
\includegraphics[width=0.98\textwidth]{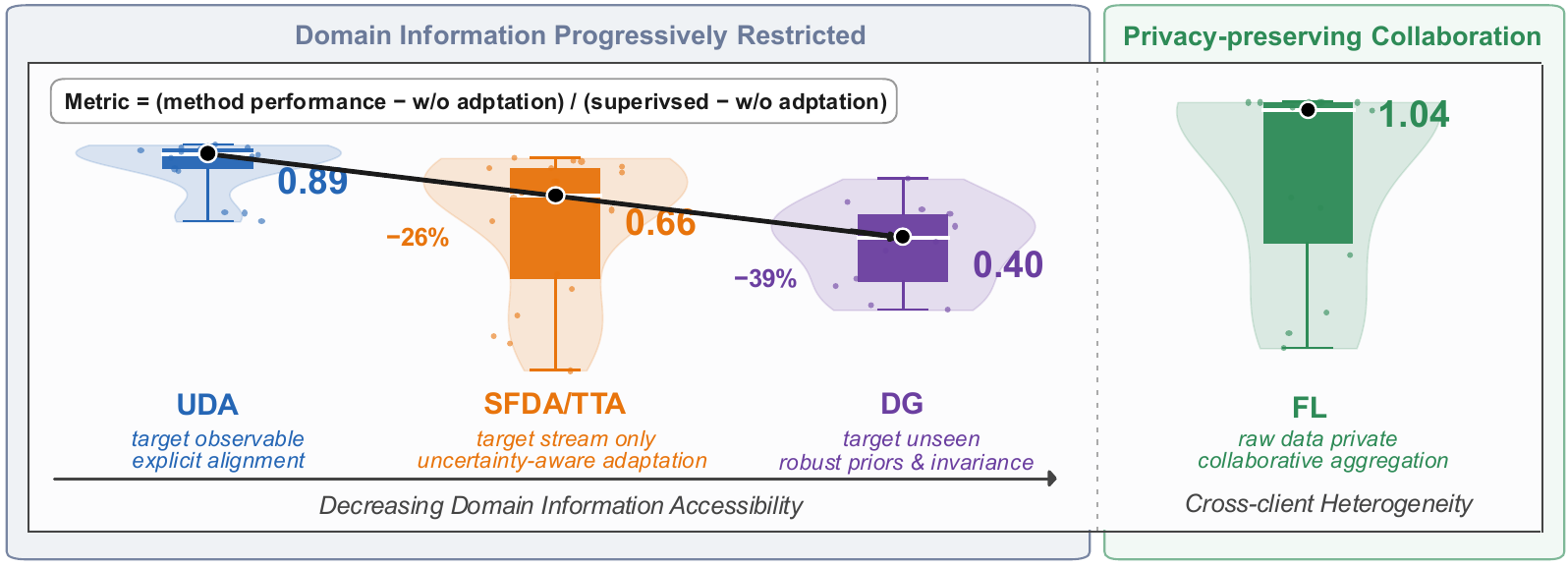}
\caption{Domain-information accessibility shapes adaptation performance. UDA, SFDA/TTA, and DG show a progressive decline in normalized recovery as domain information becomes increasingly restricted, whereas FL follows a different trend by leveraging cross-client collaboration under raw-data privacy constraints. See details in Tables~\ref{tab:cardiac_full_citations}--\ref{tab:fed_table_results} in Appendix~\ref{apdix:tables}.}
\label{fig:empirical_trends}
\end{figure*}

\section{Empirical Observations}\label{sec:Empirical Observations and Performance Trade-offs}

\subsection{Constraint-Driven Patterns Across Paradigms}
To systematically understand the evolution of method performance under different distribution shift settings, we first summarize the detailed specifications of the cross-domain benchmarks used in the literature (see Table~\ref{tab: dataset_detail}), covering cardiac, abdominal, prostate, and fundus imaging datasets. Building upon this foundation, we then aggregate experimental results from representative studies,
corresponding to Unsupervised Domain Adaptation (UDA), Source-Free Domain Adaptation and Test-Time Adaptation (SFDA/TTA), Domain Generalization (DG), and Federated Learning (FL), respectively. It is noted that these results are \textit{not} intended for strict horizontal comparison across methods for best model selection (due to systematic differences in experimental settings). Instead, we position these tables as an empirical evidence pool: rather than answering ``which method performs best'', they illuminate broader performance trends and inherent trade-offs imposed by varying information constraints.

Placing these empirical results within a unified analytical framework reveals a consistent structural phenomenon: the performance upper bound of models under distribution shift is fundamentally governed by the amount of domain information accessible during training or adaptation. As the accessibility of domain information progressively diminishes, model capability exhibits a continuous degradation trajectory, transitioning from {explicit distribution alignment} to {implicit generalization via extrapolation} (see Figure~\ref{fig:empirical_trends}).

In the UDA setting, target-domain data (albeit unlabeled) remains observable, enabling models to explicitly characterize the target distribution through feature alignment. From an optimization perspective, the distribution shift problem is thus reduced to a \textit{constrained risk minimization problem}, where the solution space is explicitly restricted by the target-domain distribution. This significantly reduces uncertainty and allows state-of-the-art methods to closely approach the fully supervised upper bound (within a 2--3\% gap). Fundamentally, this indicates that {the observability of the target domain transforms distribution shift into a controlled bias correction problem}.

In SFDA/TTA scenarios, this mechanism breaks down due to the absence of source-domain data. Conventional cross-domain alignment becomes infeasible, and models must instead rely on self-generated predictions for iterative updates. This process is better characterized as \textit{uncertainty-driven self-bootstrapping}, where the model simultaneously acts as both learner and supervisor. Within this closed loop, erroneous predictions may be progressively amplified, leading to performance degradation (approximately a 5\% gap). As such, the central challenge shifts from distribution alignment to {controlling error propagation in the absence of external reference}.

In the DG setting, the problem is further pushed to an extreme: target-domain data is entirely unavailable during training, requiring models to learn generalizable representations without any access to target distribution statistics. From a learning-theoretic standpoint, this constitutes a \textit{reference-free out-of-distribution extrapolation problem}. In the absence of explicit constraints, models must rely solely on stable structures implicitly encoded within source-domain data, which imposes stringent requirements on representation learning while simultaneously exposing its limitations (typically resulting in performance degradation of 10\% or more). This suggests that {Empirical Risk Minimization (ERM) alone is insufficient for cross-domain generalization, as it lacks the ability to disentangle invariant semantics from domain-specific perturbations}.

Notably, FL deviates from the above monotonic trend. As observed in Table~\ref{tab:fed_table_results}, certain advanced methods even surpass centralized training in terms of average performance. This phenomenon reveals a deeper mechanism: {performance is determined not only by the availability of information, but also by how it is organized and utilized}. 
While centralized ERM often prioritizes majority distributions -- leading to representation bias in long-tail domains -- the decentralized optimization in FL acts as a structural constraint that preserves local heterogeneity. By leveraging this heterogeneity as an implicit regularization signal, the model is forced to suppress site-specific artifacts and distill domain-invariant representations, ultimately enhancing representation generalizability.
Consequently, our findings suggest that distribution shift is not merely a challenge of information scarcity, but also one of structural organization and utilization.

\subsection{Algorithmic Trends: From Distribution Alignment to Representation Learning and Uncertainty Modeling}

Under this unified perspective, methodological differences across paradigms can be further abstracted into two recurring algorithmic trends, which are progressively reshaping the core technical pathways for improving cross-domain performance.

First, representation learning is becoming a central route toward robust generalization. The core challenge of distribution shift lies in the coexistence of transferable biological semantics and non-transferable domain-specific artifacts within observed data. One important direction is disentangled representation learning, which explicitly separates domain-invariant semantic or anatomical factors from domain-specific style or acquisition-related variations. By structurally decoupling what should be preserved from what should be suppressed, these methods reduce the model's reliance on spurious correlations and shift the focus from matching distributions toward {structural reconfiguration of the representation space}. Another increasingly influential direction is large-scale pre-training and foundation models. By learning high-level semantic representations from diverse data, these models improve the completeness and transferability of learned features, thereby reducing the amount of task-specific adaptation required under downstream distribution shifts. From this perspective, scaling representation learning does not directly eliminate distribution shift, but rather mitigates its impact by {enhancing the semantic completeness and transferability of learned representations}.

Second, uncertainty modeling is evolving from an auxiliary component into a central principle for reliable adaptation under limited information. In settings such as SFDA, TTA, and FL, where source data, target annotations, or centralized supervision are often unavailable, model updates increasingly depend on predictive confidence rather than ground-truth labels. Techniques such as uncertainty-guided pseudo-label filtering, confidence-weighted training, calibrated test-time adaptation, and uncertainty-aware aggregation in FL aim to prevent unreliable predictions from being amplified during adaptation. At a fundamental level, these approaches introduce an {adaptive risk control mechanism} by explicitly modeling what the model does not know under domain shift.

Taken together, these trends suggest a unified algorithmic transition: as clinical deployment constraints become increasingly stringent, method design is shifting from {explicit distribution alignment} toward a framework centered on {representation quality enhancement and uncertainty-aware risk control}.

\subsection{Implications: From Performance Optimization to Deployability-Aware Modeling}

From a broader perspective, the above analysis highlights a more fundamental issue beyond performance comparison: the core challenge in cross-domain medical image analysis lies not merely in improving model accuracy, but in achieving an effective balance between {deployability} and {performance} under realistic constraints.

Different settings inherently correspond to distinct clinical constraints: UDA assumes access to unlabeled target-domain data, SFDA/TTA reflects post-deployment online adaptation, FL captures privacy-preserving multi-institutional collaboration, and DG addresses generalization to entirely unseen environments. Within this context, observed performance gaps should not be interpreted solely as indicators of algorithmic superiority, but rather as direct consequences of differences in the {available information space} under varying conditions.
Accordingly, the central question for future research can be reformulated as:
\textit{Given practical constraints, how can models be designed to maximally exploit available information while effectively controlling the propagation of uncertainty?}

This reframing suggests a critical paradigm shift: instead of focusing solely on increasingly complex model architectures, future work should emphasize how to transform real-world constraints (e.g., data inaccessibility, unseen target domains) into effective modeling priors. Examples include leveraging heterogeneity as a regularization signal in FL, incorporating confidence regulation in uncertainty modeling, and exploiting transferable representations enabled by large-scale pre-training. From this perspective, the relationship between {constraints} and {performance} is not inherently antagonistic, but rather constitutes a form of structured resource that can be systematically exploited. Achieving principled breakthroughs within this constraint--performance interplay will remain a central research direction in cross-domain medical image analysis.

\section{Future Research Directions}\label{sec:future_directions}

\begin{figure*}
  \centering
  \includegraphics[width=0.95\textwidth]{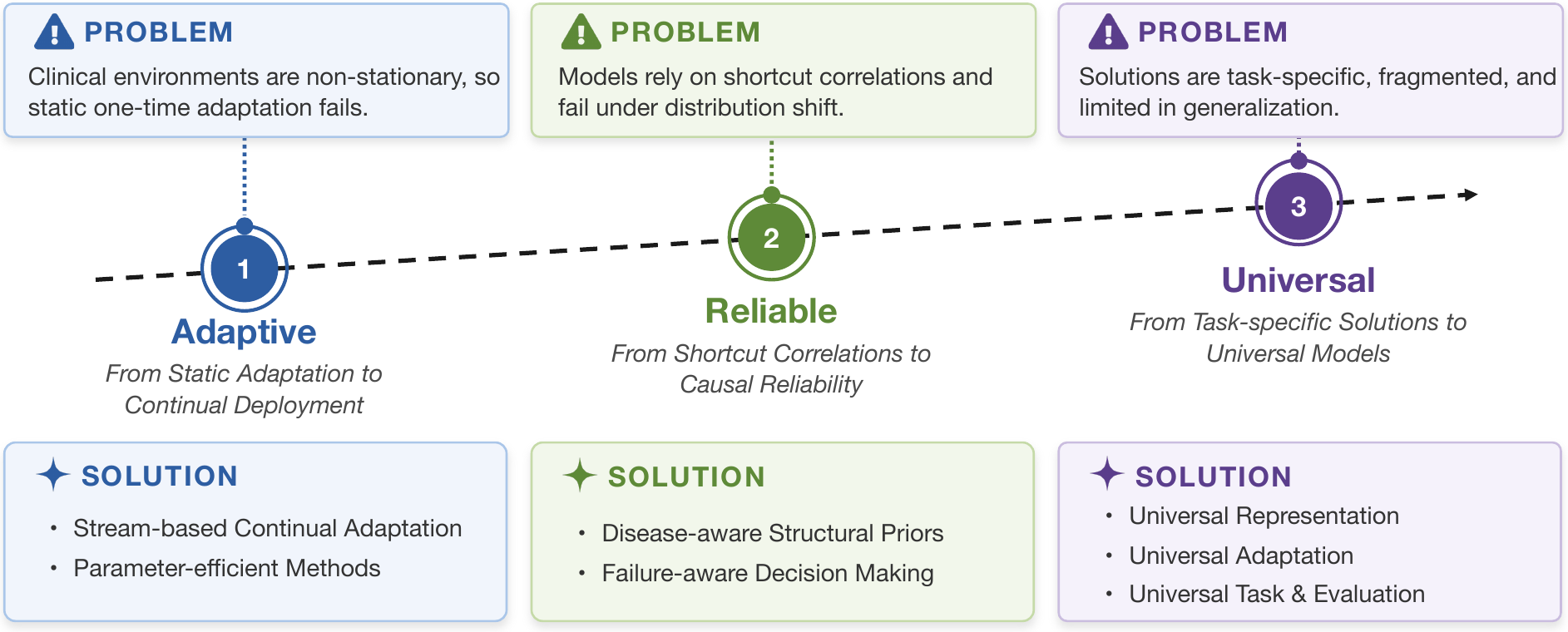}
  \caption{
  Future roadmap for distribution-shift learning in medical image analysis. We summarize three directions toward clinically deployable medical AI: adaptive learning for continual clinical deployment, reliable learning for causal and failure-aware decision making, and universal learning for reusable representations, tasks, and evaluations. They highlight a shift from static, fragmented solutions toward robust and generalizable systems under real-world distribution shifts.
  }
\label{fig:future_roadmap}
\end{figure*}

While the technical paradigms reviewed in the preceding sections have made substantial progress in mitigating distribution shifts, a persistent gap remains between experimental success and reliable clinical deployment. In real-world medical environments, distribution shifts are not isolated perturbations, but interacting consequences of evolving scanners, imaging protocols, patient populations, clinical workflows, institutional constraints, and task definitions. As a result, future research should move beyond developing task-specific remedies for individual shifts and instead pursue medical AI systems that are adaptive, reliable, and reusable across heterogeneous clinical settings (see Figure~\ref{fig:future_roadmap}).

\subsection{From Static Adaptation to Continual Deployment}
Most existing domain adaptation, domain generalization, source-free domain adaptation, and test-time adaptation methods are evaluated under relatively static assumptions, where the source and target domains are predefined and adaptation occurs as a one-time procedure. However, real clinical environments are inherently non-stationary: new scanners are installed, imaging protocols are updated, patient demographics shift, and data streams continuously accumulate across hospitals. Future models therefore need to support continual adaptation throughout the deployment lifecycle, rather than assuming a fixed target domain.

\begin{itemize}
    \item \textbf{Stream-based Continual Adaptation:}
    Clinical deployment should be treated as a continuous learning process in which models adapt to incoming data streams from evolving clinical sites while preserving previously acquired knowledge. This requires continual learning mechanisms that can update model behavior under distributional drift without catastrophic forgetting, overfitting to transient target-domain noise, or degrading performance on previously encountered domains. In medical imaging, such adaptation must also be conservative and auditable, because uncontrolled online updates may introduce clinically unsafe behavior.

    \item \textbf{Resource-aware and Parameter-efficient Adaptation:}
    Practical adaptation is constrained by limited annotation, privacy restrictions, computational cost, and edge-side deployment requirements. Instead of relying on full model fine-tuning, future methods should emphasize parameter-efficient, memory-efficient, and communication-efficient adaptation strategies. These techniques are particularly important for latency-sensitive or resource-constrained scenarios, such as intraoperative imaging, bedside ultrasound, mobile diagnostics, and hospital-local deployment. The goal is not merely faster adaptation, but low-overhead, traceable, and reversible model updating that can be integrated into real clinical workflows.
\end{itemize}

\subsection{From Shortcut Correlations to Causal Reliability}
A central reason for performance collapse under distribution shift is that deep models may rely on shortcut correlations that are predictive in the training environment but unstable across clinical sites. These shortcuts may include scanner-specific artifacts, acquisition protocols, annotation styles, demographic biases, or other non-pathological features. Future robustness research should therefore move beyond matching marginal distributions and instead encourage models to base their decisions on pathology-relevant imaging evidence, anatomical context, and calibrated uncertainty.

\begin{itemize}
    \item \textbf{Disease-aware Structural Priors:}
    Medical images are not arbitrary natural images -- they are constrained by organ topology, anatomical geometry, physiological structure, and disease-specific deformation patterns. Incorporating anatomical priors into pre-training, representation learning, and adaptation can help models distinguish clinically meaningful changes from domain-specific appearance variations. However, such priors should be disease-aware rather than rigidly invariant, because true pathology, surgery, congenital abnormalities, or disease progression may substantially alter anatomical structure. Future methods should therefore learn probabilistic structural constraints that preserve plausible anatomical organization while allowing clinically meaningful deviations.
    \item \textbf{Failure-aware  Decision Making:}
    Even with improved structural and causal modeling, some distribution shifts will remain too severe for reliable autonomous prediction. Therefore, future systems must know when they do not know. Robust uncertainty quantification, out-of-distribution detection, calibration under shift, and selective prediction are essential for identifying cases where model confidence is unreliable. In high-risk scenarios, models should trigger human review, request additional evidence, or abstain from prediction. Such failure-aware feedback loops transform robustness from a purely performance-oriented objective into a safety-critical mechanism for human-AI collaboration.
\end{itemize}

\subsection{From Task-specific Solutions to Universal Medical Foundation Models}
Current research on distribution shifts in medical image analysis is still largely organized around task-specific settings, such as adapting a segmentation model for one organ, or transferring a classifier across hospitals. Although these studies provide important insights, they often remain confined to limited clinical scenarios. A longer-term direction is to move from isolated robustness solutions toward universal medical foundation models: systems that learn reusable representations, support flexible task interfaces, and adapt efficiently across modalities, anatomical structures, institutions, and patient populations.

\begin{itemize}
    \item \textbf{Universal Representation:}
    Future models should learn representations that generalize across heterogeneous medical data sources, including MRI, CT, ultrasound, pathology, electrocardiograms, laboratory measurements, clinical reports, and longitudinal electronic health records. Instead of treating each modality as an isolated problem, universal medical models should capture shared patterns while preserving modality-specific evidence. Such representations can provide a common foundation for cross-domain transfer when target-domain labels are expensive to obtain.

    \item \textbf{Universal Adaptation:}
    Universal foundation models should not eliminate the need for local adaptation. Instead, they should make adaptation more efficient, safer, and more reliable. In clinical deployment, even a broadly pre-trained model must be calibrated to local scanners, protocols, patient populations, and workflow requirements. Future research should therefore combine universal pre-training with parameter-efficient adaptation, test-time calibration, continual learning, and source-free adaptation. The goal is to build models that possess a strong general medical prior while still allowing lightweight institution-specific refinement under privacy and resource constraints.

    \item \textbf{Universal Task:}
    Most existing models are designed with fixed task-specific heads, such as a segmentation decoder, a classification layer, or a detection module. This design limits their transferability when the clinical question changes. Future systems should instead support flexible task interfaces, allowing users to specify tasks through natural language prompts, anatomical queries, visual prompts, structured clinical templates, or interactive feedback. A universal interface would enable the same model to support segmentation, detection, classification, report generation, image retrieval, visual question answering, and decision support, without requiring a separately engineered architecture for every task.

    \item \textbf{Universal Evaluation:}
    A universal model also requires universal evaluation. Future benchmarks should move beyond testing performance on a single task or a single target domain, and instead evaluate models across multiple axes of shift, including unseen institutions, modalities, anatomical regions, diseases, demographic subgroups, and annotation protocols. Evaluation should also include calibration, uncertainty estimation, failure detection, selective prediction, and human-in-the-loop behavior. Without such broad evaluation, a model may appear universal only because it performs well on a narrow collection of retrospective benchmarks.
\end{itemize}

\section{Conclusion}
In this survey, we provide a comprehensive examination of how deep learning models can be adapted to address distribution shifts in MedIA. Rather than relying on a static technical taxonomy, we explicitly align core adaptation strategies --- Joint Training, Federated Learning, Fine-tuning, and Domain Generalization --- with key clinical constraints, including data accessibility, privacy concerns, and collaborative protocols. Building on this categorization, our empirical analysis reveals how increasingly restricted domain information reshapes model design, driving a shift from explicit data alignment toward uncertainty-aware modeling and feature disentanglement. By linking these emerging algorithmic trends with real-world MedIA constraints, we highlight that model performance is fundamentally bounded by domain information accessibility, calling for a shift from isolated benchmark optimization toward deployability-aware modeling under realistic constraints. 
Ultimately, this survey provides a principled perspective on this paradigm shift, positioning real-world constraints as structured signals for model design and offering a roadmap toward robust, deployable MedIA systems.

\section*{CRediT authorship contribution statement}
\noindent \textbf{Zixian Su}: Conceptualization, Methodology, Investigation, Formal analysis, Data curation, Visualization, Validation, Writing -- original draft \& review \& editing.  
\textbf{Jingwei Guo}: Conceptualization, Formal analysis, Visualization, Validation, Writing -- original draft \& review \& editing.  
\textbf{Xi Yang}: Writing -- review \& editing, Funding acquisition.
\textbf{Qiufeng Wang}: Writing -- review \& editing.
\textbf{Frans Coenen}: Writing -- review \& editing.
\textbf{Amir Hussain}: Writing -- review \& editing.
\textbf{Kaizhu Huang}: Writing -- review \& editing, Funding acquisition.

\section*{Appendix A. Empirial Results}\label{apdix:tables}
\noindent Tables~\ref{tab:cardiac_full_citations}--\ref{tab:fed_table_results} provide the detailed quantitative results supporting the empirical analysis summarized in Section~\ref{sec:Empirical Observations and Performance Trade-offs}.
These tables cover representative UDA, SFDA/TTA, SDG, and FL settings, and are intended to reveal broad cross-paradigm performance trends.

\clearpage
\newpage

\begin{table*}[!t]
\centering
\renewcommand{\arraystretch}{1.2}
\setlength\tabcolsep{2.2pt} 
\caption{Quantitative performance of \textbf{Unsupervised Domain Adaptation (UDA)} methods for Cardiac MRI$\rightarrow$CT segmentation. Results are categorized by backbone architectures. Symbols denote the original sources of results: * \cite{su2023mind}, $\dagger$ \cite{cai2025style}, $\ddagger$ \cite{p1xie2022unsupervised}, $\S$ \cite{7zhao2022uda}, $**$ \cite{liu2023structure}, and $\dagger\dagger$ \cite{zheng2024dual}.}
\label{tab:cardiac_full_citations}
\resizebox{0.81\textwidth}{!}{%
\begin{tabular}{l|l|ccccc|ccccc}
\hline
\multirow{2}{*}{Backbone} & \multirow{2}{*}{Method} & \multicolumn{5}{c|}{Dice (\%) $\uparrow$} & \multicolumn{5}{c}{ASD (voxel) $\downarrow$} \\ \cline{3-12}
 & & AA & LAC & LVC & MYO & Avg & AA & LAC & LVC & MYO & Avg \\ \hline

\multirow{7}{*}{DeepLabV2-ResNet50} 
 & Supervised* & 88.49 & 87.71 & 90.69 & 83.33 & 87.55 & 3.64 & 3.19 & 2.03 & 2.14 & 2.75 \\
 & W/o adaptation* & 30.75 & 26.25 & 10.10 & 1.36 & 17.12 & 26.92 & 18.45 & 16.75 & 28.78 & 22.73 \\ \cline{2-12}
 & CycleGAN* \cite{zhu2017unpaired} [ICCV 17'] & 72.58 & 71.88 & 50.45 & 31.52 & 56.61 & 12.50 & 14.76 & 9.08 & 8.48 & 11.21 \\
 & SynSeg-Net* \cite{synseg} [TMI 18'] & 66.99 & 62.11 & 47.20 & 38.00 & 53.58 & 12.85 & 8.19 & 7.31 & 10.22 & 9.64 \\
 & CyCADA* \cite{hoffman2018cycada} [ICML 18'] & 68.18 & 64.23 & 48.90 & 44.89 & 56.55 & 9.69 & 10.36 & 9.41 & 11.77 & 10.34 \\
 & DSAN \cite{han2021deep} [TMI 22'] & 79.92 & 84.76 & 82.77 & 66.52 & 78.50 & 7.68 & 6.65 & 3.77 & 5.59 & 5.92 \\
 & GLUA \cite{su2023mind} [JBHI 23'] & 87.18 & 85.47 & 87.03 & 70.28 & 82.49 & 4.65 & 3.59 & 3.50 & 3.09 & 3.71 \\ \hline

\multirow{8}{*}{DeepLabV2-ResNet101} 
 & Supervised$^\dagger$ & 92.00 & 91.60 & 93.00 & 88.20 & 91.20 & 1.50 & 3.50 & 1.80 & 2.90 & 2.40 \\
 & W/o adaptation$^\dagger$  & 0.10 & 4.50 & 25.70 & 2.80 & 8.20 & 51.00 & 42.50 & 14.70 & 10.30 & 29.60 \\ \cline{2-12}
 & UESM \cite{bian2020uncertainty} [MedIA 20'] & 84.15 & 88.30 & 84.32 & 71.42 & 82.05 & 3.87 & 3.49 & 3.81 & 3.70 & 3.71 \\
 & MPSCL \cite{p6liu2022margin} [JBHI 22'] & 90.26 & 87.08 & 86.45 & 72.51 & 84.08 & 3.47 & 3.16 & 2.85 & 3.41 & 3.47 \\
 & SECASA \cite{p10feng2023unsupervised} [AAAI 23'] & 83.80 & 85.20 & 82.90 & 71.70 & 80.90 & 9.60 & 4.20 & 3.90 & 3.90 & 5.40 \\
 & SCCDM \cite{cai2024symmetric} [ICASSP 24'] & 90.90 & 85.50 & 87.60 & 75.70 & 84.90 & 4.30 & 3.60 & 3.70 & 3.30 & 3.70 \\
 & SMEDL \cite{cai2025style} [MedIA 25'] & 88.30 & 88.20 & 86.70 & 71.10 & 83.60 & 4.30 & 2.70 & 2.50 & 3.40 & 3.20 \\ \hline

\multirow{2}{*}{ResNet-101} 
 & Supervised$^\ddagger$ & 95.76 & 92.99 & 89.75 & 84.33 & 90.71 & - & - & - & - & - \\
\cline{2-12}
 & DLaST \cite{p1xie2022unsupervised} [TMI 22'] &  89.92 & 90.09 & 86.13 & 67.59 & 83.44 & - & - & - & - & - \\ \hline
\multirow{5}{*}{U-Net} 
 & Supervised$^\S$ & 85.50 & 88.60 & 88.10 & 84.20 & 86.60 & 2.10 & 8.90 & 4.10 & 2.70 & 4.40 \\
 & W/o adaptation$^\S$ & 37.40 & 26.80 & 2.50 & 3.60 & 17.60 & 42.60 & 34.60 & - & 29.20 & - \\ \cline{2-12}
 & MT-UDA \cite{5zhao2021mt} [MICCAI 21'] & 73.10 & 82.10 & 72.80 & 61.90 & 72.50 & 22.70 & 12.20 & 11.20 & 8.20 & 13.60 \\
 & LE-UDA \cite{7zhao2022uda} [TMI 22'] & 72.90 & 83.70 & 74.60 & 62.10 & 73.30 & 23.0 & 5.80 & 6.60 & 6.60 & 10.40 \\
 \hline

\multirow{4}{*}{TransUNet} 
 & Supervised** & 91.60 & 90.90 & 90.10 & 85.40 & 89.50 & 1.90 & 1.20 & 1.60 & 1.30 & 1.50 \\
 & W/o adaptation** & 40.80 & 35.40 & 32.20 & 45.70 & 38.50 & 10.20 & 16.20 & 14.90 & 19.60 & 15.20 \\ \cline{2-12}
 & FSUDA \cite{8liu2023reducing} [AAAI 23'] & 86.80 & 87.50 & 84.60 & 82.40 & 85.30 & 1.60 & 2.50 & 3.20 & 3.10 & 2.60 \\
 & FSUDA-v2 \cite{liu2023structure} [TMI 23'] & 88.20 & 88.90 & 85.20 & 82.20 & 86.10 & 1.50 & 2.60 & 2.50 & 2.90 & 2.40 \\ \hline

\multirow{3}{*}{3D U-Net} 
 & Supervised$^{\dagger\dagger}$  & 94.30 & 91.40 & 91.80 & 86.40 & 91.00 & 1.65 & 2.43 & 2.01 & 1.86 & 1.99 \\
 & W/o adaptation$^{\dagger\dagger}$ & 56.40 & 72.40 & 53.80 & 42.70 & 56.30 & 21.84 & 14.56 & 28.85 & 24.66 & 22.47 \\ \cline{2-12}
 & DDSP \cite{zheng2024dual} [MedIA 24'] & 93.30 & 90.90 & 90.00 & 81.90 & 89.00 & 1.89 & 2.59 & 2.93 & 2.86 & 2.56 \\ \hline

\multirow{10}{*}{\begin{tabular}[c]{@{}l@{}}Custom \\ Arch.\end{tabular}} 
 & SIFA-v1 \cite{chen2019synergistic} [AAAI 19'] & 81.10 & 76.40 & 75.70 & 58.70 & 73.00 & 10.60 & 7.40 & 6.70 & 7.80 & 8.10 \\
 & SIFA-v2 \cite{chen2020unsupervised} [TMI 20'] & 81.30 & 79.50 & 73.80 & 61.60 & 74.10 & 7.90 & 6.20 & 5.50 & 8.50 & 7.00 \\
 & DSFN \cite{zou2020unsupervised} [IJCAI 20'] & 84.70 & 76.90 & 79.10 & 62.40 & 75.80 & - & - & - & - & - \\
 & ICMSC \cite{zeng2021semantic} [MICCAI 21'] & 85.60 & 86.40 & 84.30 & 72.40 & 82.20 & 2.40 & 3.30 & 3.40 & 3.20 & 3.10 \\
 & SASAN \cite{tomar2021self} [TMI 21'] & 82.00 & 76.00 & 82.00 & 72.00 & 78.00 & 4.14 & 8.30 & 3.50 & 3.30 & 4.90 \\
 & DDA-GAN \cite{DDAGAN} [MedIA 21'] & 68.30 & 75.70 & 78.50 & 77.80 & 75.10 & 6.50 & 4.80 & 5.40 & 5.20 & 5.50 \\
 & UMDA \cite{liu2021automated} [MedIA 21'] & 89.20 & 82.70 & 82.60 & 66.20 & 80.20 & 6.70 & 3.60 & 4.50 & 3.00 & 4.00 \\
 & CUDA \cite{p4du2021constraint} [JBHI 21'] & 87.20 & 88.50 & 83.00 & 72.80 & 82.90 & 7.03 & 2.80 & 5.20 & 6.80 & 5.50 \\
  & FPL+ \cite{wu2024fpl+} [TMI 24'] & 73.84 & 80.19 & 76.24 & 64.54 & 73.70 & 3.90 & 1.89 & 2.34 & 2.33 & 2.61 \\
 & MamUC-MISp \cite{feng2025mamba} [KBS 25'] & 83.79 & 77.13 & 84.87 & 79.67 & 81.33 & 1.69 & 2.49 & 1.99 & 2.41 & 2.15 \\ \hline
\end{tabular}
}
\end{table*}

\begin{table*}[!t]
\centering
\renewcommand{\arraystretch}{1.2}
\setlength\tabcolsep{2pt} 
\caption{Quantitative performance of \textbf{Source-Free Domain Adaptation (SFDA)} and \textbf{Test-Time Adaptation (TTA)} methods on the multi-organ abdominal (CT$\rightarrow$MRI) segmentation. Symbols indicate results from: * \cite{wang2023fvp}, $\dagger$ \cite{yin2025ddfp}, $\ddagger$ \cite{zhang2025enhancing}, and $\S$ \cite{stan2024unsupervised}.}
\label{tab:abdominal_comprehensive_full}
\resizebox{0.81\textwidth}{!}{
\begin{tabular}{l|l|ccccc|ccccc}
\hline
\multirow{2}{*}{Backbone} & \multirow{2}{*}{Method} & \multicolumn{5}{c|}{Dice (\%) $\uparrow$} & \multicolumn{5}{c}{ASD/ASSD $\downarrow$} \\ \cline{3-12} 
 & & Liver & R.Kid & L.Kid & Spleen & Avg & Liver & R.Kid & L.Kid & Spleen & Avg \\ \hline
\multirow{5}{*}{U-Net} & Supervised$^\dagger$ & 95.55 & 95.32 & 94.70 & 93.46 & 94.76 & 0.69 & 0.90 & 0.63 & 1.09 & 0.83 \\
 & W/o adaptation$^\dagger$ & 56.40 & 86.55 & 84.64 & 41.18 & 67.19 & 2.65 & 0.91 & 0.64 & 4.53 & 2.18 \\ \cline{2-12}
  & SFUDA\_AOS\cite{SFDA8} [KBS 22'] & {88.40} & {89.10} & {86.40} & {91.10} & {88.80} & {0.33} & 0.14 & 0.26 & 0.14 & 0.22 \\
 & ProContra$^\dagger$\cite{TTA12} [MICCAI 23'] & {79.33} & {91.32} & {88.24} & {76.41} & {83.82} & {0.33} & 3.19 & 3.81 & {1.76} & {2.27} \\
 & TT-SFUDA$^\dagger$\cite{vs2022target} [MIDL 24'] & 72.84 & 78.06 & 85.61 & 46.90 & 70.85 & 1.96 & {2.66} & {0.57} & 4.89 & 2.52 \\
 & DDFP$^\dagger$ [KBS 25'] & {90.53} & {92.06} & {92.63} & {84.26} & {89.87} & {0.83} & {0.34} &{0.46} & {4.51} & {1.54} \\ \hline
\multirow{7}{*}{DeepLabv3 (ResNet-50)} & Supervised$^\dagger$ & 93.64 & 95.20 & 93.71 & 92.94 & 93.87 & 0.49 & 0.14 & 0.29 & 0.32 & 0.31 \\
 & W/o adaptation$^\dagger$ & 76.14 & 86.95 & 77.40 & 62.14 & 75.66 & 2.17 & 1.27 & 1.35 & 2.15 & 1.74 \\ \cline{2-12}
 & DPL$^\dagger$\cite{SFDA5} [MICCAI 21'] & {87.75} & 78.60 & 57.79 & {77.33} & {75.37} & {1.41} & 3.20 & {2.27} & {1.61} & {2.12} \\
 & CBMT$^\dagger$\cite{tang2023source} [MICCAI 23'] & {84.31} & 53.11 & 76.19 & 76.52 & 72.53 & {1.72} & 3.06 & 6.60 & 2.34 & 3.43 \\
 & FSM* \cite{SFDA7} [MedIA 22'] & 63.20 & 85.40 & 79.60 & 50.80 & 69.75 & 4.77 & {2.55} & 1.72 & 6.76 & 3.95 \\
 & FVP\cite{wang2023fvp} [TMI 23'] & 64.80 & {87.60} & {80.30} & 60.50 & 73.30 & 4.48 & {2.10} & {1.54} & 6.15 & 3.57 \\
 & DDFP\cite{yin2025ddfp} [KBS 25'] & 78.06 & {89.27} & {87.47} & {85.22} & {85.01} & 1.87 & 2.56 & {0.78} & {1.31} & {1.63} \\ \hline
\multirow{14}{*}{DeepLabv3+ (MobileNetV2)} & Supervised$^{\ddagger}$ & 89.55 & 88.32 & 87.18 & 84.32 & 87.34 & 0.55 & 0.74 & 1.04 & 1.97 & 1.08 \\
 & W/o adaptation$^{\ddagger}$ & 62.59 & 66.50 & 49.55 & 44.68 & 55.83 & 4.40 & 2.01 & 3.73 & 7.11 & 4.31 \\ \cline{2-12}
  & Vanilla PL$^{\ddagger}$ & 67.40 & 69.97 & 56.67 & 56.59 & 62.66 & 2.54 & 1.99 & 3.23 & 4.72 & 3.12 \\
 & AdaEnt$^{\ddagger}$\cite{bateson2020source} [MICCAI 20'] & 83.94 & 84.08 & 78.61 & 78.73 & 81.34 & 1.95 & {1.21} & 1.74 & 3.04 & 1.99 \\
  & DPL$^{\ddagger}$\cite{SFDA5} [MICCAI 21'] & 79.62 & 82.11 & 58.67 & 75.57 & 73.99 & 2.76 & 2.44 & 2.78 & 3.65 & 2.91 \\
 & AdaMI$^{\ddagger}$\cite{TTA9} [MedIA 22'] & {84.74} & 83.08 & {82.04} & {80.74} & {82.65} & 1.84 & {1.17} & {1.11} & {2.32} & {1.61} \\
 & APL$^{\ddagger}$\cite{li2022adaptive} [ICASSP 22'] & 83.13 & 81.44 & 75.12 & 76.79 & 79.12 & 1.83 & 1.22 & 1.65 & 3.07 & 1.94 \\
 & FSM$^{\ddagger}$\cite{SFDA7} [MedIA 22'] & 83.14 & 83.65 & 73.76 & 76.01 & 79.14 & 1.95 & 1.35 & 1.74 & 2.52 & 1.89 \\
 & ProSFDA$^{\ddagger}$\cite{TTA14} [arxiv 22']& 80.45 & 77.97 & 72.16 & 77.36 & 76.99 & {1.13} & 1.46 & 1.80 & 2.46 & {1.71} \\
 & ProtoSFDA$^{\ddagger}$\cite{TTA12} [MICCAI 23'] & 84.20 & 82.42 & 78.50 & 80.10 & 81.30 & 1.82 & 2.10 & 1.26 & 2.62 & 1.95 \\
  & TT-SFDA$^{\ddagger}$\cite{vs2022target} [MIDL 24'] & 79.25 & 82.01 & 69.68 & 74.29 & 76.31 & {1.22} & 2.14 & 1.53 & 3.36 & 2.06 \\
   & RMS\cite{zhang2025enhancing} [MedIA 25'] & 85.01 & {85.18} & 79.15 & 81.01 & 82.58 & 1.76 & 1.35 & {1.03} & 2.68 & {1.71} \\ 
  \hline
  \multirow{3}{*}{DeepLabv3 (VGG-16)} 
&Supervised$^\S$   & 92.00 & 91.10 & 80.60 & 85.70 & 87.30 & 1.30 & 2.00 & 1.50 & 1.30 & 1.50 \\
  & W/o adaptation$^\S$  & 48.90 & 50.90 & 65.30 & 65.70 & 57.70 & 4.50 & 12.30 & 6.80 & 4.50 & 7.00 \\
  & SFS\cite{stan2024unsupervised} [MedIA 24'] & 86.30 & 88.00 & 85.10 & 74.90 & 83.50 & 4.50 & 1.60 & 2.20 & 18.20 & 6.60 \\\hline
\end{tabular}
}
\end{table*}

\begin{table*}[!t]
\centering
\renewcommand{\arraystretch}{1.2}
\setlength\tabcolsep{2pt} 
\caption{Quantitative performance of \textbf{Single-source Domain Generalization (SDG)} methods for prostate segmentation in cross-site scenarios. Models are trained on a single source site and evaluated on all remaining target sites. Results are grouped by backbone architectures. Symbols denote the original sources of results: * \cite{wei2024prompting}, $\dagger$ \cite{oh2025fiesta}, and $\ddagger$ \cite{yang2025domain}.}
\label{tab:prostate_cross_site}
\resizebox{0.8112\textwidth}{!}{
\begin{tabular}{l|l|cccccc|cccc}
\hline
\multirow{2}{*}{Backbone} & \multirow{2}{*}{Methods} & \multicolumn{6}{c|}{Site-wise Dice (\%) ($\uparrow$)} & \multicolumn{4}{c}{Averaged Performance} \\ \cline{3-12} 
 & & A & B & C & D & E & F & Dice ($\uparrow$) & Jaccard ($\uparrow$) & HD95 ($\downarrow$) & ASD ($\downarrow$) \\ \hline
\multirow{11}{*}{U-Net (EfficientNet-b2)} & Supervised$^\dagger$ & 83.75 & 84.78 & 84.92 & 84.98 & 86.68 & 84.92 & 85.01 & 74.31 & 5.37 & 2.59 \\
 & W/o Adaptation$^\dagger$ & 71.81 & 65.56 & 43.98 & 71.97 & 48.39 & 37.82 & 56.59 & 41.82 & 40.29 & 27.87 \\\cline{2-12}
 & Cutout$^\dagger$ \cite{Cutout} [arxiv 17'] & 78.36 & 69.08 & 63.45 & 66.39 & 61.88 & 60.19 & 66.56 & 53.54 & 20.55 & 13.16\\
 & RSC$^\dagger$ \cite{RSC} [ECCV 20'] & 72.81 & 70.18 & 49.18 & 74.11 & 54.73 & 43.69 & 60.78 & 46.37 & 30.73 & 19.68 \\
 & MixStyle$^\dagger$ \cite{mixstyle} [ICLR 21'] & 73.24 & 58.06 & 44.75 & 66.78 & 49.81 & 49.73 & 57.06 & 43.31 & 34.08 & 23.12 \\
 & AdvBias$^\dagger$\cite{advbias} [MICCAI 20'] & 78.15 & 62.24 & 54.73 & 72.65 & 53.14 & 51.00 & 61.98 & 48.56 & 27.15 & 17.59 \\
 & RandConv$^\dagger$\cite{randconv} [ICLR 21'] & 77.28 & 60.77 & 53.54 & 66.21 & 52.12 & 36.52 & 57.74 & 44.11 & 34.47 & 22.24 \\
 & CSDG$^\dagger$\cite{SDG7} [TMI 22'] & 82.14 & 67.21 & 59.11 & 73.16 & 67.38 & 73.23 & 70.37 & 59.96 & 15.45 & 9.83 \\
 & SLAug$^\dagger$\cite{SDG6} [AAAI 23] & 81.47 & 65.19 & 52.69 & 76.89 & 68.02 & 72.66 & 69.49 & 59.44 & 16.82 & 10.19\\
 & FIESTA\cite{oh2025fiesta} [TNNLS 25'] & 83.02 & 70.42 & 62.06 & 77.31 & 66.17 & 74.79 & 72.30 & 62.38 & 13.57 & 7.84 \\
 & EIR-SDG \cite{niu2025eir} [ACM MM 25'] & 81.40 & 67.99 & 68.36 & 77.82 & 67.70 & 75.58 & 72.98 & - & - & - \\ \hline
\multirow{2}{*}{SAM-b} & W/o Adaptation* & 84.42 & 79.79 & 64.83 & 83.49 & 80.50 & 80.18 & 78.87 & - & - & - \\
&DAPSAM\cite{wei2024prompting} [MICCAI 24'] & 86.34 & 81.05 & 70.81 & 85.28 & 82.91 & 81.48 & 81.31 & - & - & - \\ \hline
\multirow{3}{*}{MaskGen-Former} & Supervised$^\ddagger$ & 85.86&87.03&87.72&86.80&85.98&86.73&87.03& - & - & - \\

&W/o Adaptation$^\ddagger$ & 73.02 & 62.21 & 55.94 & 71.58 & 47.73 & 38.55 & 63.40 & - & - & - \\

&DG-DDM-Seg\cite{yang2025domain} [TMI 25'] & 82.65 & 73.24 & 66.60 & 75.55 & 71.35 & 74.37 & 78.14 & - & - & - \\ \hline
SAM-b \& ResNet-34 & CollaSU-SDG \cite{liu2025perturbating} [AAAI 25'] & 84.76 & 83.21 & 70.58 & 85.46 & 81.55 & 82.20 & 81.29 & - & - & - \\ \hline
\end{tabular}
}
\end{table*}

\begin{table*}[!t]
\centering
\renewcommand{\arraystretch}{1.2}
\setlength\tabcolsep{2.6pt} 
\caption{Quantitative performance comparison of \textbf{Federated Learning (FL)} methods for retinal fundus segmentation across six clinical centers. Results are obtained using the U-Net backbone architecture. Symbols indicate the sources of results: * \cite{chen2024fedevi} and $\dagger$ \cite{feddata6}.}
\label{tab:fed_table_results}
\resizebox{0.98\textwidth}{!}{
\begin{tabular}{l|cccccccc|cccccccc}
\hline
\multirow{2}{*}{Method} & \multicolumn{8}{c|}{Dice (\%) $\uparrow$} & \multicolumn{8}{c}{HD95 $\downarrow$} \\ \cline{2-17} 
 & C1 & C2 & C3 & C4 & C5 & C6 & Avg$\uparrow$ & Std$\downarrow$ & C1 & C2 & C3 & C4 & C5 & C6 & Avg$\downarrow$ & Std$\downarrow$ \\ \hline
Centralized$^{\dagger}$ & 91.61 & 87.60 & 87.58 & 90.22 & 85.10 & 91.79 & 88.98 & - & - & - & - & - & - & - & - & - \\  
FedAvg*\cite{fedavg} [AISTATS 17'] & 88.23 & 73.57 & 90.60 & 92.03 & 90.84 & 92.21 & 87.91 & 7.19 & 27.38 & 33.92 & 8.84 & 6.14 & 7.11 & 5.34 & 14.79 & 12.58 \\  \hline
FedProx*\cite{fedprox} [MLSys 20']& 91.48 & 77.79 & 90.85 & 92.10 & 91.02 & 92.29 & 89.25 & 5.66 & 16.61 & 27.28 & 8.53 & 5.96 & 7.09 & 5.29 & 11.79 & 8.80 \\
FedDG*\cite{MDG3} [CVPR 21']& 91.33 & 82.91 & 91.33 & 92.24 & 90.09 & 91.60 & 89.92 & 3.52 & 21.06 & 17.79 & 8.55 & 5.90 & 7.71 & 5.87 & 11.15 & 6.59 \\
{FedProto*}\cite{tan2022fedproto} [AAAI 22'] & 87.73 & 73.05 & 91.11 & 92.25 & 90.66 & 91.97 & 87.79 & 7.42 & 28.35 & 33.85 & 8.22 & 5.56 & 7.29 & 5.56 & 14.80 & 13.01 \\
FedSAM*\cite{qu2022generalized} [ICML 22'] & 89.51 & 75.87 & 91.01 & 92.36 & 90.70 & 91.76 & 88.54 & 6.30 & 20.96 & 27.54 & 8.28 & 5.54 & 7.04 & 5.44 & 12.47 & 9.55 \\
FedBN*\cite{li2021fedbn} [ICLR 21']& 89.35 & 82.99 & 92.10 & 92.17 & 89.36 & 91.94 & 89.65 & 3.62 & 21.45 & 18.52 & 7.42 & 5.62 & 9.94 & 5.52 & 11.41 & 7.33 \\
FedSM\cite{feddata6} [CVPR 22'] & 91.32 & 87.69 & 88.65 & 90.41 & 84.83 & 91.95 & 89.14 &- & - & - & - & - & - & - & - & - \\
FedBR*\cite{guo2023fedbr} [ICML 23'] & 87.09 & 74.17 & 90.00 & 89.36 & 89.03 & 91.74 & 86.90 & 6.47 & 24.35 & 30.72 & 9.70 & 10.53 & 9.18 & 5.87 & 15.06 & 10.14 \\
FedLAW*\cite{li2023revisiting} [ICML 23'] & 91.09 & 80.12 & 91.24 & 91.50 & 90.51 & 91.83 & 89.38 & 4.56 & 17.05 & 20.33 & 8.53 & 7.47 & 7.20 & 5.62 & 11.03 & 6.15 \\
FedCE*\cite{jiang2023fair} [CVPR23']& 88.50 & 74.93 & 90.85 & 92.48 & 90.97 & 91.84 & 88.26 & 6.68 & 25.43 & 27.56 & 9.23 & 5.57 & 7.11 & 5.50 & 13.40 & 10.41 \\
FedGA*\cite{zhang2023federated} [CVPR 23']& 89.20 & 81.94 & 89.92 & 89.78 & 89.08 & 91.67 & 88.60 & 3.42 & 22.42 & 19.45 & 9.05 & 7.97 & 7.46 & 5.64 & 12.00 & 7.16 \\
L-DAWA*\cite{rehman2023dawa}[ICCV 23'] & 90.85 & 79.33 & 90.14 & 91.33 & 91.03 & 92.06 & 89.12 & 4.84 & 20.37 & 21.10 & 9.83 & 6.94 & 7.21 & 5.43 & 11.81 & 7.09 \\
FedUAA*\cite{wang2023federated} [MICCAI 23'] & 92.54 & 79.99 & 90.51 & 91.83 & 90.67 & 92.37 & 89.65 & 4.81 & 13.59 & 20.59 & 8.96 & 6.24 & 6.84 & 5.10 & 10.22 & 6.02 \\
FedEvi\cite{chen2024fedevi} [MICCAI 24'] & {93.30} & {87.44} & 90.32 & 90.83 & 89.76 & {92.48} & {90.69} &{2.09} & {11.21} & {11.62} & 8.98 & 6.62 & 6.89 & {5.23} & {8.43} & {2.64} \\
\hline
\end{tabular}
}
\end{table*}

\section*{}
\newpage
\clearpage

\bibliographystyle{cas-model2-names}
\bibliography{survey}

\end{document}